\documentclass[journal]{IEEEtran}
\usepackage{amsfonts,amssymb,amsmath,bm}
\usepackage{graphicx,subfigure}
\usepackage{cite,color}
\usepackage{booktabs}
\usepackage{threeparttable}
\usepackage{mathtools}
\usepackage{verbatim}

\begin{document}



\title{Dynamic Optimization For Heterogeneous Powered Wireless Multimedia Sensor Networks With Correlated Sources and Network Coding}

%
%

 \maketitle
\begin{abstract}
The energy consumption in wireless multimedia sensor networks (WMSN) is much greater
than that in traditional wireless sensor networks. Thus, it is a huge challenge to
remain the perpetual operation for WMSN.
  In this paper, we propose a new heterogeneous energy supply model for WMSN through the coexistence of renewable energy and electricity grid.
We address to cross-layer optimization for the multiple multicast with distributed source coding and intra-session network coding in heterogeneous powered wireless multimedia sensor networks (HPWMSN) with correlated sources.
The aim is to achieve the optimal reconstruct distortion at sinks  and the minimal cost of purchasing electricity from electricity grid. Based on the Lyapunov drift-plus-penalty with perturbation technique and dual decomposition technique, we propose a fully distributed dynamic cross-layer algorithm, including multicast routing, source rate control, network coding, session scheduling and energy management, only requiring knowledge of the instantaneous system state.
The explicit trade-off between the optimization objective and queue backlog is theoretically proven. Finally, the simulation results verify the theoretic claims.
\end{abstract}

\begin{IEEEkeywords}
Wireless multimedia sensor networks, heterogeneous energy, cross-layer optimization,  Lyapunov optimization, network coding, distributed source coding.
\end{IEEEkeywords}

\section{Introduction}

\IEEEPARstart{W}{ireless} Multimedia Sensor Networks (WMSN) are composed of a large number of
heterogeneous multimedia sensors,
interconnected through wireless medium to  capture video and audio
streams, still images, and scalar sensor data, and
transmits them to the sink(s) through the wireless manner.
WMSN  has a lot of potential new applications, for example, multimedia surveillance, traffic avoidance and control, health monitoring, environment monitoring, automated manufacturing processes, and industrial process control \cite{Akyildiz2007}.


Similar to wireless sensor networks (WSN), the sensor nodes in WMSN are traditionally battery-operated, where the battery is usually limited and non-rechargeable. As the multimedia applications in WMSN require high transmission rate and extensive processing, the energy consumption of WMSN is much greater than that of WSN. Moreover, the sensors may be placed in remote and dangerous areas, the replacement of sensors is unrealistic and impossible. Hence, the network lifetime is an important issue for the operation of the most applications in WMSN, and an efficient power management should be proposed to extend the network lifetime.
Recently, a large number of works focus on the energy efficiency  for WMSN with the limited battery capacity \cite{Ehsan2012}\cite{Mo2011}. Although the energy efficiency techniques increases the sensor's lifetime, the network lifetime still has an upper bound due to the limited battery capacity. In order to solve this problem, the renewable energy technique is applied to power the sensors via solar panels or wind-powered generators \cite{Raghunathan2006}. However, the renewable energy technique exits some shortcoming, such as low recharging rate and time-varying profile of energy replenishment process, which can not guarantee to provide the perpetual operation for WMSN. As the electricity grid can provide persistent power input, the coexistence of renewable energy and electricity grid, called Heterogeneous Power (HP) is expected to achieve the infinite network lifetime\cite{Gong2013}.

Since the sensors in MWSN are always densely deployed in the most application scenarios, and the sensors generally have large monitoring range. For example,  the nearby camera sensors's monitoring range may be overlapped, and the multimedia information flows generated by them have a certain degree of spatial correlation and redundancy \cite{Dai2009}. By removing the correlation and redundancy of the multimedia data, the amount of data transmitted in the network can be reduced, thereby reducing the energy consumption and improving the energy efficiency for MWSN.
Distributed source coding (DSC) is one of the techniques that can remove the correlation and redundancy of multiple correlated sources, without communication between the sources.
By modeling the correlation between multiple sources at the decoder side,
 DSC is able to shift the computation burden in sources to the joint decoder \cite{Xiong2004}\cite{Puri2006}. Thus, DSC provides appropriate frameworks for WMSN with complexity-constrained sensors and powerful sink(s). It is more important to use DSC in HPWMSN, since DSC enables multiple correlated sources to trade energy resources among them.

Network coding (NC) can combine various traffic flows or packets  into a single packet in  the intermediate node via  simple algebraic operations, and then forwards it through one or more outgoing links\cite{Ahlswede2000}. Thus, NC has the potential of achieving substantial throughput and power efficiency gains in wireless networks.
NC applied in WMSN can save a lot of energy consumption, and then prolong the lifetime of  WMSN.

In this paper, we address a discrete-time stochastic cross-layer optimization problem with joint DSC and NC for HPWMSN to achieve the high energy efficiency and the desirable network performce. The key contribution are summarized as follows:
\begin{itemize}
  \item Since most multimedia applications in WMSN require extensive data processing, we propose the overall energy consumption model to include multiple energy consumptions due to data processing, transmission and reception. In order to obtain potentially infinite lifetime, we propose the new energy heterogenous supply model through the coexistence of renewable energy and electricity grid.  Sensors in  WMSN can be powered by renewable energy, or electricity grid or both.
  \item We address to cross-layer optimization for the multiple multicast with DSC and NC in HPWMSN with correlated sources.
We formulate a discrete-time stochastic cross-layer optimization problem
with the goal of  minimizing the time-average utility of reconstruct distortion at sinks and minimizing the cost of purchasing energy from electricity grid subject to network strong stability, link capacity, and data/energy availability constraints.
This will involves all the layer of the network, such as the transmission layer, the network layer, especially the physical layer and medium access layer.

  \item Based the Lyapunov drift-plus-penalty with perturbation technique, we transfer the stochastic optimization problem into a deterministic optimization problem. By Lagrange dual decomposition, the problem is decoupled into five subproblems, including energy harvesting and battery charging/discharging problem, source rate control problem, distortion control problem, information/physical flow rate and power control problem, and session scheduling. Finally, a fully distributed iterative algorithm is designed.

  \item
We analyze the performance of the proposed distributed algorithm, and show that a control parameter $V$  enables an explicit trade-off between the average objective value and queue backlog. Specifically, our algorithm can achieve a time average objective value that is within ${\cal {O}}(1/V)$ of the optimal objective for any $V >0$, while ensuring that the average queue backlog is ${\cal  {O}}(V)$.
Finally, the simulation results verify the theoretic claims.
\end{itemize}

Throughout this paper, we use the following notations.
We denote the probability Pr$(A)$ of
an event $A$. For a random variable
$X$, $\mathbb{E}[X]$ represents its expected value, $\mathbb{E}[X|{A}]$ represents its expected
value conditioned on event $A$.
$\textbf{1}_{A}$ is the indicator function for an event $A$, it is
1 if $A$ occurs and is 0 otherwise. ${{{[x]}^ + } = \max (x,0)}$.

The remainder of the paper is organized as follows.
The related works are given in Section II.
In Section III, we give the system model and problem formulation.
In Section IV, we present
the distributed cross-layer optimization algorithm.
In Section V, we present the performance analysis of
our proposed algorithm.
Simulation results are given in Section VI.
Concluding remarks are provided in Section VI.

\section{Related Works}

\subsection{Distributed Source Coding}

Using Slepian-Wolf model and a joint entropy coding model, Cristescu et al. in \cite{Cristescu2004} propose
a minimum cost of data transmission optimization problem in WSNs, where the sources are spatially
correlated. Wang et al. in \cite{Wang2008} propose a methodology for cross-layer optimization
between routing and DSC in WSNs. In order to meet the data transmission latency and QoS requirements, Khoshroo et al. in \cite{Khoshroo} jointly consider the resource allocation, channel coding, and DSC.
Cheng et al. in \cite{Cheng2013} proposed a hierarchical DSC coding scheme to completely remove inter-node data correlation and redundancy.
Wang et al. in \cite{Wang2013} jointly consider the optimization of Slepian-Wolf (SW) source coding and transmission rates to obtain the maximum source rate over a Gaussian multiple access channel.
However, these works do not consider the link capacity constraint, which may cause congestion and deteriorate the network performance. Hence,
 Yuen et al. in \cite{Yuen2008} introduce the fixed link capacity constraint. They propose a distributed algorithm of joint rate allocation and transmission structure in WSN with correlated data.
 In practice, the link capacity depends on the power allocated to the link and the channel state.
 Ramamoorthy in \cite{Ramamoorthy2011}  introduce the unfixed link capacity constraints, and consider a minimum cost optimization problem in WSN with multiple correlated sources. He et al. in \cite{He2012} firstly focus on  the network lifetime maximization that jointly considers routing, power control, and link-layer random access.

 As far as we know, almost all works usually ignore the energy consumption of the data processing. However, in some applications, the energy consumption of data processing may be comparable to the energy consumption of data transmission. Hence,
Tapparello et al. in \cite{Tapparello2014} consider the joint energy allocation for communication module and processing module together in the multihop WSN scenario. They show that through applying DSC, the sensor with sufficient energy can ease the transmission requirements of correlated data on nearby node with low energy. This characteristic of DSC is very suitable for energy-harvesting networks.
In \cite{Stankovic}, Stankovic et al. in \cite{Stankovic} apply DSC to improve the network performance, consider the multimedia multicast over heterogeneous wireless-wireline networks, and propose a network-aware cross-layer optimization problem. In  \cite{Puri2006}, Puri et al. apply distributed video coding to model the correlation between multiple sources at the decoder side together with channel codes. To the end, the encoder is simple, and the computational complexity is shifted to the decoder. Hence DSC is a promising  technique for WMSN.

\subsection{Network Coding}

The network coding techniques have been widely used in WSN. Chachulski et al. \cite{Chachulski2007} exploit the broadcast nature of wireless transmission and propose a random network coding scheme. Li et al. in \cite{Li2012} consider multirate multicast with network coding over wireless video networks, to optimize the video quality and network performance. Lin et al. in \cite{Lin2013} propose an approximate optimization scheme to
jointly optimize the link scheduling, rate control, and flow allocation problems for multicast with intraflow network coding in multirate multichannel wireless mesh networks.
Ho et al. \cite{Ho2006} design a random linear network coding (RLNC) for wireless transmission. Chen et al. in \cite{Chen2012} study the multicast flow control based network coding for wired networks and wireless networks. Rajawat et al. in \cite{Rajawat2011} consider the random network coding for slotted wireless multihop networks. These works show that NC can obtain throughput gain and save a lot of energy and thus prolong the network lifetime.
However,
these works usually assume the data transmission over link is free of interference, which weakens the applicability of NC technique.
 Hence, some works address to NC in interference-limited networks.
Xi et al. in \cite{Xi2010} propose a framework for the minimum-cost optimization problem in interference-limited wireless networks, and jointly consider the power allocation, network coding and multicast.
NC and DSC are jointly considered in \cite{Ho2009} and \cite{Cordeschi2013}. Given a  time-varying networks,  Ho et al. in \cite{Ho2009} propose a dynamic algorithms for the problem of multiple multicast sessions with intra-session network coding. \cite{Cordeschi2013} jointly consider the adaptive DSC, channel coding,
NC and power control with  QoS constraint for Co-Channel Interference
(CCI)-limited wireless networks. However, they do not take
the energy constraint into consideration, which is one of the
most important constraint in WMSN.

\section{SYSTEM MODEL}
\label{sec:model}

We consider a general interconnected multi-hop WSN that operates over time slots $t \in {\cal T}=\left\{ {0,1,2, \ldots } \right\}$. WSN is modeled by a direct graph
${\cal G} = \left\{ {{\cal N},{\cal L}} \right\}$.
${\cal N}  = {{\cal N}_H} \cup {{\cal N}_G} \cup {{\cal N}_M} = \left\{ {1,2,3, \ldots ,N} \right\}$ denotes the set of sensor nodes in the network, ${\cal N}_H$ is the set of EH nodes powered by renewable energy, ${\cal N}_G$ is the set of EG nodes powered by electricity grid, and ${\cal N}_M$ is the set of ME nodes powered by both renewable energy and electricity grid, respectively. ${\cal N}_s=\{1,2,\ldots,N_s\} \subset {\cal N}$ denotes the set of all source nodes which measure the information source(s), therein, we note that the sources are continuous and correlated. ${\cal N}_d =\{1,2,\ldots,N_d\}\subset {\cal N}$ denotes the set of all sink nodes. Let ${\cal F} =\{1,2,\ldots, F\}$ denote the set of  sessions in the network, and we assume the sessions are multicast, i.e., for $\forall f \in {\cal F}$, the session $f$ has at least one source and one sink. Therefore, we define
${\cal N}_s^f=\left\{ {1,2,3, \ldots ,N_s^f} \right\}$ and ${\cal N}_d^f=\left\{ {1,2,3, \ldots ,N_d^f} \right\}$ as the set of session $f'$s sources and sinks, respectively,
${\cal N}_s^f \subset {{\cal N}_s}$, ${\cal N}_s^d \subset {{\cal N}_d}$.
The source node  transmits the data to the corresponding sink node through multi-hop routing.
We use ${\cal O}\left( n \right)$ to denote the set of nodes $m$ with $(n, b) \in \cal L$,
and ${\cal I}\left( n \right)$ to denote the set of nodes $a$ with $(a, n) \in \cal L$. ${\cal L}{\rm{ = }}\left\{ {(n,b),n,b \in {\cal N}} \right\}\,b \in {\cal O}\left( n \right)$ represents the set of communication links.

\begin{figure}[t]
\includegraphics[width=3.5in]{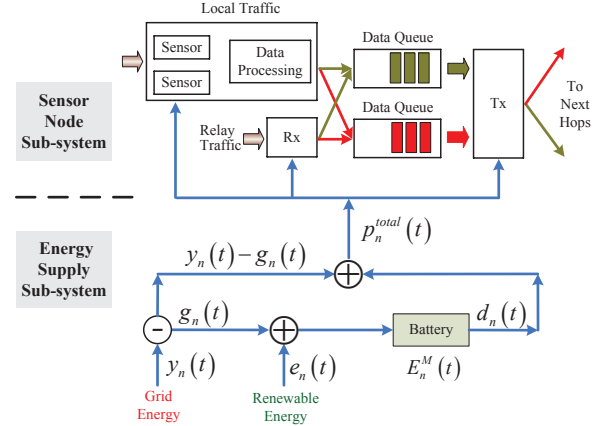}
\caption{Node Composition.}
\label{fig_Node composition}
\end{figure}

Fig. \ref{fig_Node composition} describes the composition of a single node system, which is divided into two sub-systems, i.e., sensor node sub-system and energy supply sub-system. Sensor node sub-system includes data sensing/processing and data transmission/receiption. Their energy consumption is given in the subsection \ref{energy_supply_sub-system1}.
The energy supply sub-system is detailed described in the subsection \ref{energy_supply_sub-system2}.

\subsection{Network and Coding Model}

In our network, in order to achieve the optimal throughout when transmit multiple multicast sessions, we apply the network coding technology. Note that the sink nodes only receive fractional sessions, combing different sessions together brings difficulty for data reconstruction, hence, we assume the network coding is limited within each session.

Let $x _{{nb}}^f\left( t \right)$  denote the physical flow rate of the session $f$ over link $\left( {n,b} \right)$, and $\tilde{x}_{nb}^{fsd}(t)$ denote the information flow rate of the session $f$ over link $\left( {n,b} \right)$ from source $s\in {\cal N}_s^f$ to sink $d\in {\cal N}_d^f$, where $b \in {\cal O}\left( n \right)$. Via network coding, the flows for different destinations of a multicast session can be coded together so that they can share the network capacity, while the actual physical flow over each link need only to be the maximum of the individual destinations' flows, i.e.,
\begin{equation}\label{networkcoding}
\tilde{x} _{nb}^{fsd}\left( t \right) \le x_{nb}^f\left( t \right),\;f \in {\cal F},\;s \in {\cal N}_s^f,\;d \in {\cal N}_d^f,\;b \in {\cal O}\left( n \right)
\end{equation}
\subsection{Distributed Lossy Coding}

For continuous sources, we consider the lossy source coding which allows a reconstruction distortion.
In our model, the distributed lossy coding can be described as the follows.

At every time slot $t$, for each session $f$, the node $n\in {\cal N}_s^f $ measures data from the environment and compresses it before transmission. We assume the node $n\in {\cal N}_s^f $ multicasts data to all the sinks of session $f$ at rate $r_n^f(t)$ with distortion ${D_n^f\left( t \right)}$. Consider any subset
$\forall {\cal \underline{N}}_s^f \subseteq {\cal N}_s^f$,  for each node  $n\in {\cal \underline{N}}_s^f$ ,we have:
\begin{equation}\label{source rate constraint1}
  \sum\limits_{n \in {\cal \underline{N}}_s^f} {r_n^f}(t)  \ge H\left( {{\cal \underline{N}}_s^f|{\cal N}_s^f-{\cal \underline{N}}_s^f} \right)-\log((2{\pi}e)^{|{\cal \underline{N}}_s^f|}\prod\limits_{n \in {\cal \underline{N}}_s^f} {D_n^f\left( t \right)})
\end{equation}
where $H\left( {{\cal \underline{N}}_s^f|{\cal N}_s^f-{\cal \underline{N}}_s^f} \right)$ represents the entropy of ${\cal \underline{N}}_s^f$ conditioned on ${\cal N}_s^f-{\cal \underline{N}}_s^f$.

We also assume that
\begin{equation}\label{source_rate_constraint}
 0 \le {r_n^f}(t) \le R_{\max }{\rm{,}}\forall f \in {\cal F},n\in {\cal N}_s^f
\end{equation}
and
\begin{equation}\label{distortion_constraint}
 D_{\min} \le {D_n^f}(t) \le D_{\max }{\rm{,}}\forall f \in {\cal F},n\in {\cal N}_s^f
\end{equation}

\subsection{Random Access and Data Transmission}

We assume that the links in the network may interfere with each other when they  transmit data simultaneously, 
we also assume each node cannot transmit or receive at the same time, and define ${q_n}\left( t \right)$ as the transmission probability of node $n$ at slot $t$, with
\begin{equation}\label{rand access}
  0 \le {q_{n}}\left( t \right) \le 1, \forall n \in {\cal N}.
\end{equation}
When node $n$ is selected to transmits data at slot $t$, the probability of choosing outgoing link $\left( {n,b} \right), b \in {\cal O}\left( n \right)$ is ${q_{nb}}\left( t \right)$, with $\sum\limits_{b \in {\cal O}\left( n \right)} {{q_{nb}}\left( t \right)}  = {q_n}\left( t \right)$ .

Then, we introduce the transmission probability matrix ${\bm \jmath}\left( t \right) = \left\{ {{q_{nb}}\left( t \right),\left( {n,b} \right) \in {\cal L}|0 \le {q_{nb}}\left( t \right) \le 1} \right\}$ . Summary, the link $(n,b)$ can transmit data successfully at time slot $t$ with probability

$
{\alpha _{nb}}\left( t \right) = {q_{nb}}\left( t \right)\underbrace {(\prod\limits_{a \in {\cal I}\left( n \right)} {(1 - {q_{an}}\left( t \right)} ))}_{\scriptstyle\;\rm{The}\,\rm{probability}\;\hfill\atop
{\scriptstyle\;\rm{of}\,\rm{no}\,\,\rm{acception}\hfill\atop
\scriptstyle\;\;\;\quad \rm{of}\,\rm{n}\hfill}}(\underbrace {1 - \sum\limits_{c \in {\cal O}\left( b \right)} {{q_{bc}}\left( t \right)} }_{\scriptstyle\;\rm{The}\,\rm{probability}\;\rm{of}\,\hfill\atop
{\scriptstyle\;\rm{no}\,\,\rm{transmission}\hfill\atop
\scriptstyle\;\;\;\quad \rm{of}\,\rm{b}\hfill}})$,\\

We define ${\bm P}\left( t \right){\rm{ = }}\left( {{p_{nb}^T}\left( t \right),\left( {n,b} \right) \in {\cal L}} \right)$  as the power allocation matrix for data transmission at slot $t$, where ${p_{nb}^T}(t)$  is the power allocated to link $\left( {n,b} \right)$, and each node $n$ satisfies
\begin{equation}\label{energy constraint}
0 \le \sum\limits_{b \in {\cal O}\left( n \right)} {{p_{nb}^T}(t)}  \le {P_n^{\max }},
\end{equation}

Next, we use ${\gamma _{nb}}(t)$ to denote the signal to interference plus noise ratio (SINR) of link $\left( {n,b} \right)$, which can be calculated by the following:
\begin{equation}\label{sinr}
{\gamma _{nb}}(t) = \frac{{{S _{nb}^C(t)}{p_{nb}^T}\left( t \right)}}{{{N_{nb}} + \sum\limits_{a \in {\cal I}_{n,b}}\sum\limits_{\left( {a,c} \right) \in {\cal L}} {{S _{ab}^C(t)}{p_{ac}^T}\left( t \right){q_{ac}}\left( t \right)} }},
\end{equation}
where ${\cal I}_{n,b}$ represents the set of nodes whose transmission can interfere the transmission over link $(n,b)$, ${S _{nb}^C(t)}$  represents the link fading coefficient from $n$ to $b$, and ${N_{nb}}$ represents the noise which is assumed to be constant.We assume that
${S _{nb}^C}(t)$ may be time varying and i.i.d. at every slot.
Denote ${\bm S}^C(t)=\{{S}_{nb}^C(t), (n, b) \in \cal L\}$ as
the network channel state matrix, taking non-negative values from a
finite but arbitrarily large set ${\cal S}^C$.

The link capacity is defined as the following:
\begin{equation}\label{link capacity}
{C_{nb}}\left(t \right) =
BW{\alpha _{nb}(t)}\log \left( {{\gamma _{nb}(t)}} \right)
\end{equation}
where $BW$ is the bandwidth.

 Because of the total rates of all sessions cannot exceed the link capacity, so, the following constraints must be met:
\begin{equation}\label{link_rate constraint}
0 \le \sum\limits_{f \in {\cal F}} {x _{{nb}}^f\left( t \right)} \le C _{nb}(t), \forall n \in {\cal N}, \forall b \in {\cal O}\left( n \right)
\end{equation}
Without loss of generality, we assume that for all time over all links under any power
allocation matrix and any channel state, there exists some finite constant $X_{\max}$
such that
\begin{equation}\label{Xmax}
0 \le \sum\limits_{f \in {\cal F}} {x _{{nb}}^f\left( t \right)} \leq X_{\max}
\end{equation}
\subsection{Energy Consumption Model}\label{energy_supply_sub-system1}
The energy consumption model is divided into three parts:data sensening/processing, data transmission and data reception. Therein, we define  ${\tilde{P}_f^S}$ and ${\tilde P_n^R}$ as the cost of per data sensing and reception, respectively.

When node $n\in {\cal N}_s^f$ achieves the compressed data rate at $r_n^f(t)$ , we assume the energy consumed is $\tilde{P}_f^Sr_n^f(t)$, which is linear. When node $n$ receives session $f \in \cal F$ from its neighbor node $a \in {\cal I}_n$ at rate $x _{{an}}^f\left( t \right)$, the energy required is $P_n^Rx _{{an}}^f\left( t \right)$. Above all, the total energy consumption $p_n^{Total}\left( t \right)$ of node $n$ at slot $t$ is:
\begin{eqnarray}\label{EnergyConsumption}
p_n^{Total}\left( t \right) &\buildrel\Delta \over =&\sum\limits_{f \in {\cal F}}\textbf{1}_{n \in {{\cal N}_s^f}} {\tilde{P}_f^S}{r_n^f}\left( t \right)  + \sum\limits_{b \in {\cal O}\left( n \right)} {p_{nb}^T\left( t \right)} \nonumber\\
&+&\tilde P_n^R \sum\limits_{a \in {\cal I}\left( n \right)} {\sum\limits_{f \in {\cal F}} {x _{{an}}^f\left( t \right)} }
\end{eqnarray}

\subsection{Energy Supply Model}\label{energy_supply_sub-system2}

First, we describe the energy supply model of ME node shown in Fig.\ref{fig_Node composition}.
Each ME node is equipped with a battery.
As depicted in Fig.\ref{fig_Node composition}, the harvested energy $e_n(t)$ at time $t$ for
ME node $n$ is stored in the battery.
On the other hand, the energy supplied by the electricity grid at time $t$ for
ME node $n$ is denoted with $y_n(t)$.
Some of this grid energy is used to charge the battery at a charging rate $g_n(t)$,
while the remaining energy $y_n(t)-g_n(t)$  is used to supply the operation of ME node $n$.
In addition, some energy from the battery is
discharged at a discharging rate $d_n(t)$ to supplement the energy
drawn from the electricity grid so as to meet the energy demand of
the node.
Thus, the total energy consumption $p_n^{Total}\left( t \right)$ of a
ME node $n$ at slot $t$
is determined by the energy supplied by the electricity grid, and the energy discharged from the battery, therefore expressed as
\[p_n^{Total}\left( t \right) = {y_n}\left( t \right) - {g_n}\left( t \right) + {d_n}\left( t \right)\]
which is equivalent to
\begin{equation}\label{EG_energysupply}
{y_n}\left( t \right) \buildrel\Delta \over ={g_n}\left( t \right)-{d_n}\left( t \right)+ p_n^{Total}\left( t \right),\forall n \in  {\mathcal N}_G \cup {\mathcal N}_M.
\end{equation}

\subsubsection{Energy queue}
We assume each ME-node $n$ knows its own current energy
availability ${{E_n^M}}\left( t \right)\;$ denoting the energy queue size for $\,n \in {\cal N}_M\,$ at time slot $t$.
We define ${\bm E}_M\left( t \right) = \left( {{{E_n^M}}\left( t \right),n \in {\cal N}_M} \right)$ over time slots $t \in \cal T$ as the vector of the energy queue sizes.
For ME node $n$,  the energy queuing dynamic equation is
\begin{equation}\label{EMqueue}
E_n^M\left( {t + 1} \right) = E_n^M\left( t \right) + {{e_n}\left( t \right) + {g_n}\left( t \right)} - {d_n}\left( t \right)
\end{equation}
with $E_n^M\left( 0 \right) =0$.
In any time slot $t$, the total energy consumption at ME node $n$ must satisfy the following \emph{energy-availability} constraint:
\begin{equation}\label{EMenergy_availability}
 {{E_n^M}}\left( t \right)\geq {d_n}\left( t \right),\;\; \forall n \in {{\cal N}_M}.
\end{equation}

\subsubsection{Two special cases}

In fact, both EH node and EG node are two special cases of  ME node.

For EH node, ${y_n}\left( t \right) = 0, \; {g_n}\left( t \right) = 0$, and so $p_n^{Total}\left( t \right) = {d_n}\left( t \right)$. Thus, the energy queuing dynamic equation at EH node $n$ is changed from \eqref{EMqueue} into:
\begin{equation}\label{EHqueue}
{{E_n^H}}\left( {t + 1} \right) = {{E_n^H}}\left( t \right) +  {e_n}\left( t \right) - p_n^{Total}\left( t \right),\;\;  n \in {{\cal N}_H},
\end{equation}
with $E_n^H\left( 0 \right) =0$.
The \emph{energy-availability} constraint is changed from \eqref{EMenergy_availability} into:
\begin{equation}\label{EHenergy_availability}
{{E_n^H}}\left( t \right)\geq p_n^{Total}\left( t \right),\;\; \forall n \in {{\cal N}_H}.
\end{equation}

For EG node, ${e_n}\left( t \right) = 0$.
Thus, the energy queuing dynamic equation at EG node $n$ is changed from \eqref{EMqueue} into:
\begin{equation}\label{EGqueue}
{{E_n^G}}\left( {t + 1} \right) = {{E_n^G}}\left( t \right) +{g_n}\left( t \right)-{d_n}\left( t \right),\;\;  n \in {{\cal N}_G}
\end{equation}
with $E_n^G\left( 0 \right) =0$.
The \emph{energy-availability} constraint is changed from \eqref{EMenergy_availability} into:
\begin{equation}\label{EGenergy_availability}
 {{E_n^G}}\left( t \right)\geq {d_n}\left( t \right),\;\; \forall n \in {{\cal N}_G}.
\end{equation}
We define ${\bm E}_H\left( t \right) = \left( {{{E_n^H}}\left( t \right),n \in {\cal N}_H} \right)$ and ${\bm E}_G\left( t \right) = \left( {{{E_n^G}}\left( t \right),n \in {\cal N}_G} \right)$ over time slots $t \in {\cal T}$ as the vector of the energy queue sizes for all EH nodes and that for all EG nodes, respectively.

\subsubsection{Some constraints}
We assume the available amount of harvesting energy at slot $t$  is ${h_n}\left( t \right)$
 with ${h_n}\left( t \right) \le {h_{\max }}$ for all $t$.
The amount of actually harvested energy ${e_n}\left( t \right)$ at slot $t$, should satisfy
 \begin{equation}\label{harvesting_decision}
0 \leq {e_n}\left( t \right) \leq  {h_n}(t), \forall n \in {{{\cal N}_H} \cup {{\cal N}_M}},
 \end{equation}
where ${h_n}\left( t \right)$  is randomly varying over time slots in an i.i.d. fashion according
to a potentially unknown distribution and taking non-negative values from a
finite but arbitrarily large set ${\cal S}^H$.
We define the harvestable energy vector ${{\bm S}^H}(t) = \left( {{h_n}\left( t \right), n \in {{{\cal N}_H} \cup {{\cal N}_M}} }\right)$, called the harvestable energy state.

The charging rate ${g_n}\left( t \right)\;$  of the battery of node $n$ at slot $t$ should satisfy:
 \begin{equation}\label{charging_rate constraint}
 0\le {g_n}\left( t \right) \le g_n^{\max }, \forall n \in {{{\cal N}_G} \cup {{\cal N}_M}},
 \end{equation}
 with some finite $g_n^{\max }$.

The discharging rate ${d_n}\left( t \right)\;$  of the battery of node $n$ at slot $t$ should satisfy:
 \begin{equation}\label{discharging_rate constraint}
 0\le {d_n}\left( t \right) \le d_n^{\max }, \forall n \in {{{\cal N}_G} \cup {{\cal N}_M}},
 \end{equation}
with some finite $d_n^{\max }$.

 the energy supplied by the electricity grid ${y_n}\left( t \right)\;$  of the node $n$ at slot $t$ should satisfy:
 \begin{equation}\label{energy supplied constraint}
 0\le {y_n}\left( t \right) \le y_n^{\max }, \forall n \in {{{\cal N}_G} \cup {{\cal N}_M}},
 \end{equation}
with some finite $y_n^{\max }$.

Moreover, at any time slot $t$, we also assume the total energy volume stored in battery on node $n \in {\cal N}_H$ is limited by the weight perturbation $\theta _{n}^{{eH}}$ introduced in section IV, the reason will be explained in \textbf{Theorem 1}, thus the following inequations must be satisfied
\begin{equation}\label{limitedcapacityeh}
{{E_n^H}}(t)+{e_n}\left( t \right)\le { \theta _{n}^{{eH}}},\forall n \in {{\cal N}_H}
\end{equation}

\subsection{Electricity Price Model}

At time slot $t$, the cost of purchasing one unit electricity from the electricity grid at node $n \in  {\mathcal N}_G \cup {\mathcal N}_M$ is characterized by the function $P_n^{G}(t)$, which may be with respect to the energy $y_n(t)$ supplied by EG and the electricity price state variable $S_n^G(t)$. We assume $S_n^G(t)$ is an i.i.d. process and takes non-negative values from a
finite but arbitrarily large set ${\cal S}^G$.
Denote ${\bm S}^{G}(t)=\{S_n^{G}(t), n \in {{{\cal N}_G} \cup {{\cal N}_M}}\}$ as the electricity price vector.
Therein, we assume that
$P_n^{G}(t)$ is a function of both $S_n^G(t)$ and $y_n(t)$, i.e.,
\begin{equation*}
P_n^{G}(t)= P_n^{G}(S_n^G(t), y_n(t))
\end{equation*}
Note that the dependence of $P_n^{G}(t)$ on $S_n^G(t)$
and $y_n(t)$ is implicit for notational convenience in the sequel.
For each $S_n^G(t)$, $P_n^{G}(t)$
is assumed to be a increasing and continuous convex function of $y_n(t)$.

\subsection{Data Queue Model}

For $f \in {\cal F}$ at node $n$, we use ${Q}_{n}^{fsd}\left( t \right)\;$ to denote the data backlog of the $f$-th session  from source
$s \in {\cal N}_s^f$ to sink $d \in {\cal N}_d^f$ at slot $t$.
We define ${{\bm{Q}}}\left( t \right) = \left( {{Q}_n^{fsd}\left( t \right),n \in {\cal N},f \in {\cal F},s \in {\cal N}_s^f,d \in {\cal N}_d^f} \right)$ over time slots $t \in \cal T$ .
Then, the data queuing dynamic equation at the network layer is
\begin{eqnarray}\label{data_queue1}
{Q}_n^{fsd}(t{\rm{ + 1}}) &=& {Q}_n^{fsd}(t) - \sum\limits_{b \in {\cal O}\left( n \right)} {\tilde{x} _{{nb}}^{fsd}\left( t \right)}\nonumber\\
&+& \sum\limits_{a \in {\cal I}\left( n \right)} {\tilde{x}_{{an}}^{fsd}\left( t \right)}
+{\textbf{1}_{f \in {{\cal F}_n}}}{r_n^f}\left( t \right).
\end{eqnarray}
with ${Q}_{n}^{fsd}\left( 0 \right) =0$. Also, the following \emph{data-availability} constraint at the network layer should be satisfied:
\begin{eqnarray}\label{Data_availability}
0&&\leq \sum\limits_{b \in {\cal O}\left( n \right)} {\tilde{x}  _{{nb}}^{fsd}\left( t \right)} \leq {{{Q}_n^{fsd}}}\left( t \right),\nonumber\\
 &&\forall n \in {{\cal N}},f \in {{\cal F}},s \in {\cal N}_s^f,d \in {\cal N}_d^f.
\end{eqnarray}



To ensure the network is strongly stable, the following inequation must be satisfied:
\begin{equation}\label{qstable}
\mathop {\lim }\limits_{T \to \infty } \frac{1}{T}\sum\limits_{t = 0}^{T - 1} {\sum\limits_{n \in {\cal N}} {\sum\limits_{f \in {\cal F}} {\sum\limits_{s \in {\cal N}_s^f} {\sum\limits_{d \in {\cal N}_d^f}  \mathbb{E}{\left\{ {Q_n^{fsd}(t)} \right\} < \infty } } } } }
\end{equation}

\subsection{Optimization Problem}\label{sec:opt}

 For each session $f$, at each node $n \in {\cal N}_s^f$, we assume the utility is
${U_n^f\left( {D_n^f\left( t \right)} \right)}$ with the corresponding distortion $D_n^f\left( t \right)$. We assume the ${U_n^f\left( \cdot \right)}$ is decreasing and concave. So, our goal
is to design a full distributed algorithm that achieves the optimal trade-off between the time-average utility of
the distortion and the time-average cost of energy consumption
in electricity grid subject to all of the
constraints described above.

Mathematically,
we will address the stochastic optimization problem \textbf{P1} as follows:
\begin{eqnarray}\label{opt}
\mathop {\mbox{maximize}}_{\{{\bm \chi}(t), t \in \cal T\}} &&\overline O = \mathop {\lim }\limits_{T \to \infty } \;\frac{1}{T}\sum\limits_{t = 0}^{T - 1} \mathbb{E}{\left\{ {O\left( t \right)} \right\}} \\
\mbox{subject to}&&
\eqref{networkcoding}-\eqref{energy constraint},
\eqref{link_rate constraint},
\eqref{Xmax},
\eqref{EG_energysupply},
\eqref{EMenergy_availability},\nonumber\\
&&\eqref{EHenergy_availability},
\eqref{EGenergy_availability},
\eqref{harvesting_decision}-\eqref{limitedcapacityeh},
\eqref{Data_availability},
\eqref{qstable}\notag
\end{eqnarray}
with
the queuing dynamics \eqref{EMqueue} for $\forall n \in {{\cal N}_M}$, \eqref{EHqueue} for $\forall n \in {{\cal N}_H}$, \eqref{EGqueue} for $\forall n \in {{\cal N}_G}$,  and \eqref{data_queue1} for $\forall n \in {\cal N}, \forall f \in {\cal F},s \in {\cal N}_s^f,d \in {\cal N}_d^f$.

${\bm \chi }(t) \buildrel \Delta \over =({\bm e}(t),{\bm r}(t),{\bm D}(t),{\bm y}(t),{\bm g}(t),{\bm d}(t),{\bm{p}}^T(t), {\bm x}(t),\tilde{{\bm x}}(t),\\{\bm q}(t))$ is the set of the optimal variables of the problem \textbf{P1}, where
${\bm e}(t)$, ${\bm r}(t)$, ${\bm D}(t)$, ${\bm y}(t)$, ${\bm g}(t)$, ${\bm d}(t)$, ${\bm{p}}^T(t)$, ${\bm x}(t)$ ,$\tilde{{\bm x}}(t)$, ${\bm q}(t)$ are the vector of ${e_n}(t)$, ${r_n^f}(t)$, ${D_n^f}(t)$, ${y_n}(t)$, ${g_n}(t)$, ${d_n}(t)$, ${p_n^T}(t)$,  ${x}_{nb}^f(t)$, $\tilde{x}_{nb}^{fsd}(t)$, $q_{nb}(t)$, respectively.
\begin{eqnarray}\label{object}
O\left( t \right) &=&\varpi_1\sum\limits_{f \in {\cal F}} \sum\limits_{n \in {\cal N}_s^f}U_n^f(D_n^f(t))\nonumber\\
&-&(1-\varpi_1)\varpi_2\sum\limits_{n \in {{{\cal N}_G} \cup {{\cal N}_M}}}{P_n^{G}(t)}{y_n(t)}\notag
\end{eqnarray}

\section{Distributed Cross-layer Optimization Algorithm}\label{sec:alg}

In this section, we assume the rand access probabilities are known a prior, so we will only determine the energy harvesting, the energy purchasing and battery dischage/charge, energy allocation, routing and scheduling decisions.

We will propose a fully distributed algorithm
which makes greedy decisions at each time slot without requiring
any statistical knowledge of the harvestable energy states, of the electricity price states and of the channel states.

\subsection{Lyapunov optimization}
First, we introduce  the weight perturbation ${{\bm \theta}  ^{{eH}}} = \left( {\theta _{n}^{{eH}},n \in {{\cal N}_H}} \right)$, ${{\bm \theta}  ^{{eM}}} = \left( {\theta _{n}^{{eM}},n \in {{\cal N}_M}} \right)$ and ${{\bm \theta}  ^{{eG}}} = \left( {\theta _{n}^{{eG}},n \in {{\cal N}_G}} \right)$.
Then we define the network state at time slot $t$ as\\
${\bm{Z}}(t) \buildrel \Delta \over =({{\bm S}^C}(t), {{\bm S}^H}(t),{{\bm S}}^G(t),{\bm{Q}}(t), {\bm{E_M}}(t),{\bm{E_H}}(t),{\bm{E_G}}(t))$.
Define the Lyapunov function as
\begin{eqnarray}\label{Lyapunov_function}
{L}(t)&=&\frac{1}{2}\sum\limits_{n \in {\cal N}} {\sum\limits_{f \in {\cal F}}\sum\limits_{s \in {\cal N}_s^f}\sum\limits_{d \in {\cal N}_d^f} {{{\left( {{Q}_n^{fsd}(t) } \right)}^2}} }\nonumber\\
&+&\frac{1}{2}\sum\limits_{n \in {{\cal N}_H}} {{{\left( {{{E_n^H}}\left( t \right) - \theta _{n}^{{eH}}} \right)}^2}}\nonumber\\
&+&\frac{1}{2}\sum\limits_{n \in {{\cal N}_M}} {{{\left( {{{E_n^M}}\left( t \right) - \theta _{n}^{{eM}}} \right)}^2}}\nonumber\\
&+&\frac{1}{2}\sum\limits_{n \in {{\cal N}_G}} {{{\left( {{{E_n^G}}\left( t \right) - \theta _{n}^{{eG}}} \right)}^2}}
\end{eqnarray}
\textbf{Remark 3.1}
From the above equation (\ref{Lyapunov_function}), we can see that when minimizing the Lyapunov function $L(t)$, we push the queue backlog towards the corresponding perturbed variable value. So, as long as we choose appropriate perturbed variables, the constraint \eqref{EMenergy_availability}, \eqref{EHenergy_availability} and \eqref{EGenergy_availability} will always be satisfied. The detailed proof will be given in the next section. Thus, we can get rid of this constraint in the sequel.

Now define the drift-plus-penalty as \\
$ {\Delta _V}\left( t \right) \buildrel \Delta \over =\mathbb{E} \left(L(t+1) - L(t)- VO(t)|{\bm{Z}}(t)\right)$,
where $V$ is a non-negative weight,  which can be tuned to control $\overline{O}$ arbitrarily close to the optimum with a corresponding tradeoff in average queue size.
Next, the upper bound of ${\Delta _V}\left( t \right)$ is given in
\begin{equation}\label{upperbound15}
 {\Delta _V}\left( t \right) \le B +\mathbb{E}\left( {\hat{{\Delta}} _V}\left( t \right)\left| {{\bm{Z}}(t)} \right.\right),
\end{equation}
where ${\hat{\Delta} _V}\left( t \right)$
is shown in \eqref{upperbound15A},
\begin{figure*}
\begin{eqnarray}
{\hat{{\Delta}} _V}\left( t \right)&=&\sum\limits_{n \in {\cal N}} {\sum\limits_{f \in {\cal F}} {\sum\limits_{s \in {\cal N}_s^f} {\sum\limits_{d \in {\cal N}_d^f} {Q_n^{fsd}(t)\left( {\sum\limits_{a \in {\cal I}(n)} {\tilde x_{an}^{fsd}(t)}  + {{\bm 1}_{n = s}}r_s^f(t) - \sum\limits_{b \in {\cal O}(n)} {\tilde x_{nb}^{fsd}(t)} } \right)} } } }  \nonumber\\
&+& {\sum\limits_{n \in {{\cal N}_H}} {\left( {{{E_n^H}}\left( t \right) - \theta _n^{{eH}}} \right)\left({e_n}\left( t \right) - p_n^{Total}\left( t \right)\right)} }\nonumber\\
&+& {\sum\limits_{n \in {{\cal N}_M}} {\left( {{{E_n^M}}\left( t \right) - \theta _n^{{eM}}} \right)\left({e_n}\left( t \right) + {g_n}\left( t \right) - {d_n}\left( t \right)\right)} }\nonumber\\
&+& {\sum\limits_{n \in {{\cal N}_G}} {\left( {{{E_n^G}}\left( t \right) - \theta _n^{{eG}}} \right)\left( {g_n}\left( t \right) - {d_n}\left( t \right)\right)} }\nonumber\\
 &-& V \left({\varpi_1\sum\limits_{f \in {\cal F}} {\sum\limits_{n \in {\cal N}_s^f} {U_n^f\left( {D_n^f\left( t \right)} \right)}  -(1-\varpi_1)\varpi_2 \sum\limits_{n \in {{\cal N}_G} \cup {{\cal N}_M}} {P_n^G\left( t \right){y_n}\left( t \right)} } }\right)\label{upperbound15A}
  \end{eqnarray}
\hrule
\end{figure*}

\begin{eqnarray}
  {B}&=& \sum\limits_{n \in {\cal N}} \sum\limits_{f \in {\cal F}}\sum\limits_{s \in {\cal N}_s^f}\sum\limits_{d \in {\cal N}_d^f}B_{{Q}}+\sum\limits_{n \in {{\cal N}_H}}{B_E^H}\notag\\
  &+&\sum\limits_{n \in {{\cal N}_M}}{B_E^M}+\sum\limits_{n \in {{\cal N}_G}}{B_E^G},\notag
\end{eqnarray}
and ${l_{\max }}$ denotes the largest number of the outgoing/incoming links that any node in the network can have.
${B_{{Q}}}{\rm{ = }}\frac{3}{2}l_{\max }^2{{X}} _{\max }^2 + \frac{1}{2}{R_{\max }^{\rm{2}}}$,
${B_E^H} =\frac{1}{2}{\left( h_{\max } \right)^2}+\frac{1}{2}{\left( {P_{n,\max }^{Total}} \right)^2}$,
${B_E^M} =\frac{1}{2}{\left( {h_{\max }+g_n^{\max}} \right)^2}+\frac{1}{2}{\left( {d_{n}^{\max}} \right)^2}$,
${B_E^G} =\frac{1}{2}{\left( {g_n^{\max}} \right)^2}+\frac{1}{2}{\left( {d_{n}^{\max}} \right)^2}$,
$P_{n,\max }^{Total} = \sum\limits_{f \in {{\cal F}}}\textbf{1}_{n \in {{\cal N}_s^f}}{\tilde P_f^S{R_{\max }}} + P_n^{\max } + \tilde P_n^R{l_{\max }}{X_{\max }}$.

After using the Lyapunov optimization, our optimization problem \eqref{opt} is changed into minimizing ${\hat{{\Delta}} _V}\left( t \right)$ i.e., \textbf{P2} as follows:
\begin{eqnarray}\label{opt1}
\mathop {\mbox{minimize}} && {\hat{{\Delta}} _V}\left( t \right) \\
\mbox{subject to}&&
\eqref{networkcoding}-\eqref{distortion_constraint},
\eqref{energy constraint},
\eqref{link_rate constraint},
\eqref{Xmax},
\eqref{EG_energysupply},\nonumber\\
&&\eqref{harvesting_decision}-\eqref{limitedcapacityeh},
\eqref{Data_availability}\notag
\end{eqnarray}

As our optimization problem \eqref{opt1} is non-convex, first, we will show that  it can be an equivalent convex problem by using the $log$ change.

First, let ${\hat p_{nb}^T}(t) = \log \left( {{p_{nb}^T}(t)} \right)$,    the definition \eqref{EnergyConsumption} of $p_n^{Total}\left( t \right)$ is changed into
\begin{eqnarray}\label{New EnergyConsumption}
p_n^{Total}\left( t \right) &\buildrel\Delta \over =&\sum\limits_{f \in {\cal F}}\textbf{1}_{n \in {{\cal N}_s^f}} {\tilde{P}_f^S}{r_n^f}\left( t \right)  + \sum\limits_{b \in {\cal O}\left( n \right)} e^{\hat{p}_{nb}^T\left( t \right)} \nonumber\\
&+&\tilde P_n^R \sum\limits_{a \in {\cal I}\left( n \right)} {\sum\limits_{f \in {\cal F}} {x _{{an}}^f\left( t \right)} }
\end{eqnarray}
the constraint \eqref{energy constraint} changes as follows
\begin{equation}\label{new energy constraint}
\sum\limits_{b \in {\cal O}\left( n \right)} e^{{\hat{p}_{nb}^T}(t)}  \le {P_n^{\max }},
\end{equation}
and the  constraint \eqref{link_rate constraint} is changed into the following:
\begin{equation}\label{new link_rate constraint}
\sum\limits_{f \in {\cal F}} {x _{{nb}}^f\left( t \right)} \le \tilde{C} _{nb}(t), \forall n \in {\cal N}, \forall b \in {\cal O}\left( n \right)
\end{equation}
where $\tilde{C} _{nb}(t)= BW{\rho _{nb}(t)}\log \left( {{{\tilde \gamma }_{nb}}\left( t \right)} \right)$, with
\begin{equation}\label{sinrtilde}
{\tilde{\gamma} _{nb}}(t) = \frac{{{S _{nb}^C(t)}e^{{\hat{p}_{nb}^T}\left( t \right)}}}{{{N_{nb}} + \sum\limits_{a \in {\cal I}_{n,b}}\sum\limits_{\left( {a,m} \right) \in {\cal L}} {{S _{ab}^C(t)}e^{{\hat{p}_{am}^T}\left( t \right)}{q_{am}}\left( t \right)} }}.
\end{equation}

Then, plugging \eqref{New EnergyConsumption} into \eqref{upperbound15A},
and rearranging all terms of the righthand side (RHS) in \eqref{upperbound15A}, the ${\hat{{\Delta}} _V}\left( t \right)$ is changed into ${\widetilde{\Delta} _V}\left( t \right)$ in \eqref{upperbound15B}.
\begin{figure*}
 \begin{eqnarray}\label{upperbound15B}
{\widetilde{\Delta} _V}\left( t \right)&=& \sum\limits_{n \in {{\cal N}_H}} {\left( {{{E_n^H}}(t) - \theta _n^{{e_H}}} \right) {e_n}(t)} + \sum\limits_{n \in {{\cal N}_M}} {\left( {{{E_n^M}}(t) - \theta _n^{{e_M}}} \right) {e_n}(t)} \nonumber\\
&+& \sum\limits_{n \in {{\cal N}_G}}\left[ \left({ E_n^G\left( t \right) - \theta _n^G} \right)({g_n}\left( t \right)-{d_n}\left( t \right))+V(1-\varpi_1)\varpi_2{P_n^{G}}(t){y_n}(t)\right]\nonumber\\
&+&\sum\limits_{n \in {{\cal N}_M}}\left[{\left({ E_n^M\left( t \right) - \theta _n^M} \right)({g_n}\left( t \right)-{d_n}\left( t \right))+V(1-\varpi_1)\varpi_2{P_n^{G}}(t){y_n}(t)}\right] \\
&-&\sum\limits_{f \in {\cal F}} {\sum\limits_{n \in {\cal N}_s^f} {\left[ {V\varpi_1U_n^f\left( {D_n^f(t)} \right) - \sum\limits_{d \in {\cal N}_d^f} {Q_n^{fnd}(t)} r_n^f(t) + {{\bm 1}_{n \in {{\cal N}_H}}}\left( {E_n^H\left( t \right) - \theta _n^{eH}} \right)\tilde P_f^Sr_n^f(t)} \right]} } \nonumber\\
 &-& \sum\limits_{n \in {\cal N}} {\sum\limits_{b \in {\cal O}(n)} {\sum\limits_{f \in {\cal F}} {\sum\limits_{s \in {\cal N}_s^f} {\sum\limits_{d \in {\cal N}_d^f} {\tilde x_{nb}^{fsd}(t)\left( {Q_n^{fsd}(t) - Q_b^{fsd}(t)} \right)} } } } } \nonumber\\
  &-&\sum\limits_{n \in {\cal N}} {\sum\limits_{b \in {\cal O}(n)} {\left[ {\sum\limits_{f \in {\cal F}} {{{\bm 1}_{b \in {{\cal N}_H}}}\left( {E_b^H\left( t \right) - \theta _b^{eH}} \right)\tilde P_b^Rx_{nb}^f\left( t \right)}  + {{\bm 1}_{n \in {{\cal N}_H}}}\left( {E_n^H\left( t \right) - \theta _n^{eH}} \right)e^{\hat{p}_{nb}^T\left( t \right)}} \right]} }\nonumber
\end{eqnarray}
\hrule
\end{figure*}

Finally, our optimization problem \textbf{P2} is changed into the equivalent problem \textbf{P3} as follows:
\begin{eqnarray}\label{opt2}
\mathop {\mbox{minimize}} && {\widetilde{\Delta} _V}\left( t \right) \\
\mbox{subject to}&&
\eqref{networkcoding}-\eqref{distortion_constraint},
\eqref{Xmax},
\eqref{EG_energysupply},
\eqref{harvesting_decision}-\eqref{limitedcapacityeh},\nonumber\\
&&\eqref{Data_availability},
\eqref{new energy constraint},
\eqref{new link_rate constraint}\nonumber
\end{eqnarray}
Above all, we have the following \textbf{Theorem 1}: Assume the random access probabilities are known a prior and the battery capacity of EH nodes have an upper bound ${{\bm \theta}  ^{{eH}}}$. By  using the $log$ change for \eqref{link capacity}, the optimization problem \textbf{P3} is convex.

\textbf{Proof}:First, we show the object function  in \eqref{opt2} is convex. Due to the function of  ${\bm e}(t)$,${\bm r}(t)$,${\bm y}(t)$,${\bm d}(t)$,${\bm x}(t)$ and $\tilde{{\bm x}}(t)$ is linear, hence ,we only have to show the function of ${\bm D}(t)$,${\bm g}(t)$ and ${\bm{p}}^T(t)$ is convex.By assuming the the battery capacity of EH nodes have an upper bound ${{\bm \theta}  ^{{eH}}}$, we have that ${{{E_n^H}}(t) - \theta _n^{{e_H}}}<0, \forall n \in {{\cal N}_H}$.
Due to the the electricity price  $P_n^{G}(t)$
is assumed to be a increasing and continuous convex function of $g_n(t)$ for each given $S_n^G(t)$, the ${U_n^f\left( \cdot \right)}$ is concave and the convexity of $e^{\hat{p}_{nb}^T(t)}$, we can get the conclusion that our object function is a convex function of $\bm {\chi}(t)$.

Second, we show the constraints in \eqref{opt2} are convex. We can easily have the constraints \eqref{networkcoding}-\eqref{distortion_constraint},
\eqref{Xmax},
\eqref{EMenergy_availability},
\eqref{EHenergy_availability},
\eqref{EGenergy_availability},
\eqref{harvesting_decision}-\eqref{limitedcapacityeh},
\eqref{Data_availability} are convex, and due to the convexity of $e^{\hat{p}_{nb}^T(t)}$, we can have \eqref{EG_energysupply},\eqref{new energy constraint} are convex.
 Next,
\begin{eqnarray}\label{Psi}
&& \log \left( {{\tilde{\gamma} _{nb}}(t)} \right)
 = \log {S _{nb}^C(t)} + \hat p_{nb}^T(t) \\
 &&- \log \left( {N_{nb} + \sum\limits_{a \in {\cal I}_{n,b}}\sum\limits_{\left( {a,m} \right) \in {\cal L}} {S _{ab}^C(t)q_{am}(t)e^ {\hat p_{am}^T(t)} } } \right).\nonumber
\end{eqnarray}
it is not difficult to prove that $\log \left( {{\tilde{\gamma} _{nb}}(t)} \right)$ is a strictly concave
function of a logarithmically transformed power vector $\bm{\hat p}^T(t)$\cite{Boyd_Convex}, so \eqref{new link_rate constraint} is convex.

\subsection{Framework of }
The framework of our algorithm is summarized in TABLE I.
\begin{table}[h]
\centering
\caption{Algorithm 1}
\begin{tabular}{|p{3in}|}
\hline
\begin{itemize}
\item [1] \; Initialization:
The perturbed variables ${{\bm \theta} ^{{e_H}}}$, ${{\bm \theta}^{{e_G}}}$, ${{\bm \theta}^{{e_M}}}$ and the penalty parameter $V$ is given.
\item [2] \; Repeat at each time slot $t \in \cal T$:
\item [3]  \quad Observe ${\bm Z} (t)$;
 \item [4]  \quad Choose the set ${\bm{\chi}^{*}(t)}$ of the optimal variables as the
optimal solution to the following optimization problem \textbf{P3}.
\item [5]  \quad Update the energy queues and data queues according to \eqref{EMqueue}, \eqref{EHqueue}, \eqref{EGqueue} and \eqref{data_queue1}, respectively.
\end{itemize}
\\
\hline
\end{tabular}
\end{table}

\subsection{Dual decomposition}

Therein, we apply the dual decomposition to solve problem \textbf{P3}.

As the object is to minimize the function of ${\bm y}(t)$, and the function is convex, so the constraint \eqref{EG_energysupply} is equivalent to the following inequation
\begin{equation}\label{new EG energysupply}
{y_n}\left( t \right) \ge {g_n}\left( t \right) - {d_n}\left( t \right) + p_n^{Total}(t).
\end{equation}
Obviously, due to the constraint \eqref{source rate constraint1} in \textbf{P3},  the variables ${\bm r}(t)$ and ${\bm D}(t)$ are coupled. And due to the constraint \eqref{new EG energysupply}, the variables ${\bm y}(t)$, ${\bm g}(t)$, ${\bm d}(t)$, ${\bm{\tilde{p}}}^T(t)$, ${\bm x}(t)$ and ${\bm r}(t)$ are coupled, we can decouple them by using the Lagrangian dual method. We introduce two dual variables ${\bm \lambda}(t)$ and ${\bm \rho}(t)$, with
${\bm \lambda} \left( t \right) = \left( {{\lambda _n}\left( t \right),\forall n \in {{\cal N}_G} \cup {{\cal N}_M}} \right)$,
${\bm \rho} \left( t \right) = \left( {\rho _m^f\left( t \right),\forall f \in {\cal F},m = 1,2, \ldots ,{2^{{N}_s^f - 1}}} \right)$.

Thus, the problem \textbf{P3} is changed into \textbf{P4}:
\begin{eqnarray}\label{opt3}
\mathop {\mbox{minimize}} && {\cal L}\left( {{\bm \chi} \left( t \right),{\bm \lambda} \left( t \right),{\bm \rho} \left( t \right)} \right) \\
\mbox{subject to}&&
\eqref{networkcoding},
\eqref{source_rate_constraint},
\eqref{distortion_constraint},
\eqref{Xmax},
\eqref{harvesting_decision}-\eqref{limitedcapacityeh},\nonumber\\
&&\eqref{Data_availability},
\eqref{new energy constraint},
\eqref{new link_rate constraint}\notag
\end{eqnarray}
where ${\cal L}\left( {{\bm \chi} \left( t \right),{\bm \lambda} \left( t \right),{\bm \rho} \left( t \right)} \right)$
is the Lagrangian function of problem \textbf{P3},
\begin{eqnarray}\label{L}
&&{\cal L}\left( {{\bm \chi} \left( t \right),{\bm \lambda} \left( t \right),{\bm \rho} \left( t \right)} \right)= {\tilde{\Delta} _V}(t)\nonumber\\
 &&+ \sum\limits_{n \in {{\cal N}_G} \cup {{\cal N}_M}} {{\lambda _n}\left( t \right)} \left( {{g_n}\left( t \right) - {d_n}\left( t \right) + p_n^{Total}(t)- {y_n}\left( t \right)} \right)\nonumber\\
 &&+ \sum\limits_{f \in {\cal F}} \sum\limits_{m = 1}^{{2^{N_s^f}} - 1} \rho _m^f\left( t \right)[ H\left( {\underline{{\cal N}}_{sm}^f|{\cal N}_s^f - \underline{{\cal N}}_{sm}^f} \right){\rm{ - }}\sum\limits_{n \in \underline{{\cal N}}_{sm}^f} {r_n^f\left( t \right)}\nonumber\\
  &&{\rm{ - }}\log ( {( {2\pi e} )}^{|\underline{{\cal N}}_{sm}^f|}\prod\limits_{n \in \underline{{\cal N}}_{sm}^f} {D_n^f( t )})]
\end{eqnarray}
Therein, we note that for each session $f$, there exits ${2^{N_s^f}} - 1$ subsets $\underline{{\cal N}}_{sm}^f$ of its source set ${\cal N}_s^f$.

Plugging \eqref{upperbound15B} into \eqref{L}, we have
 \begin{flalign}\label{L1}
&{\cal L}\left( {{\bm \chi} \left( t \right),{\bm \lambda} \left( t \right),{\bm \rho} \left( t \right)} \right)&\nonumber\\
&= \sum\limits_{n \in {{\cal N}_H}} {\left( {{{E_n^H}}(t) - \theta _n^{{e_H}}} \right) {e_n}(t)}
 + {\cal L}_1({\bm e}(t),{\bm g}(t),{\bm \lambda}(t))& \nonumber\\
&+{\cal L}_2({\bm d}(t),{\bm y}(t),{\bm \lambda}(t))+ {\cal L}_3({\bm g}(t),{\bm d}(t),{\bm y}(t),{\bm \lambda}(t))&\notag\\
&- {\cal L}_4({\bm D}(t),{\bm \lambda}(t))\nonumber- {\cal L}_5({\bm r}(t),{\bm \lambda}(t),{\bm \rho}(t)&\notag\\
&-{\cal L}_6({\bm x}(t),\tilde{{\bm x}}(t),{\bm {\hat{p}}}^{T}(t),{\bm \lambda}(t))&
\end{flalign}
where
\begin{flalign}
 & {\cal L}_1({\bm e}(t),{\bm g}(t),{\bm \lambda}(t))&\\
 &=\sum\limits_{n \in {{\cal N}_M}}[ \left({ E_n^M\left( t \right) - \theta _n^M+{\lambda _n}} \right){g_n}\left( t \right)&\notag\\
 &+\left({ E_n^M\left( t \right) - \theta _n^M} \right){e_n}\left( t \right)],&\nonumber
\end{flalign}
\begin{flalign}
 & {\cal L}_2({\bm d}(t),{\bm y}(t),{\bm \lambda}(t))&\nonumber\\
&=\sum\limits_{n \in {{{\cal N}_M}}}[\left({V(1-\varpi_1)\varpi_2{P_n^{G}}(t)-{\lambda _n}}\right){y_n}(t)&\notag\\
&-\left({ E_n^M\left( t \right) - \theta _n^M+{\lambda _n}} \right){d_n}\left( t \right)]&\nonumber\\
\end{flalign}
\begin{flalign}
 & {\cal L}_3({\bm g}(t),{\bm d}(t),{\bm y}(t),{\bm \lambda}(t))&\nonumber\\
 &=\sum\limits_{n \in {{\cal N}_G}}[ \left({ E_n^G\left( t \right) - \theta _n^G+{\lambda _n}} \right)({g_n}\left( t \right)-{d_n}\left( t \right))&\nonumber\\
&+\left({V(1-\varpi_1)\varpi_2{P_n^{G}}(t)-{\lambda _n}}\right){y_n}(t)],&
\end{flalign}
\begin{flalign}
 & {\cal L}_4({\bm D}(t),{\bm \lambda}(t))&\notag\\
 &=\sum\limits_{f \in {\cal F}} \sum\limits_{n \in {\cal N}_s^f} [ V\varpi_1U_n^f\left( {D_n^f(t)} \right)& \nonumber\\
 &+ \log \left( {D_n^f(t)} \right)\sum\limits_{m:n \in \underline{{\cal N}}_{sm}^f} {\rho _m^f\left( t \right)} ] ,&
\end{flalign}
\begin{flalign}
 & {\cal L}_5({\bm r}(t),{\bm \lambda}(t),{\bm \rho}(t))&\notag\\
 &=\sum\limits_{f \in {\cal F}} \sum\limits_{n \in {\cal N}_s^f} [ \sum\limits_{m:n \in \underline{{\cal N}}_{sm}^f} {\rho _m^f\left( t \right)}&  \nonumber\\
 &- \sum\limits_{d \in {\cal N}_d^f} {Q_n^{fnd}(t)}  + A_n(t)\tilde P_f^S ]  r_n^f(t),&
\end{flalign}
with
\begin{equation}\label{Anb}
A_{n}(t) \buildrel \Delta \over = {\textbf{1}_{n \in {{\cal N}_H}}}\left( {{{E_n^H}}(t) - \theta _n^{{eH}}} \right) - {\textbf{1}_{n \in {{{\cal N}_G} \cup {{\cal N}_M}}}}{\lambda}_n
\end{equation}
\begin{flalign}
 &{\cal L}_6({\bm x}(t),\tilde{{\bm x}}(t),{\bm {\hat{p}}}^{T}(t),{\bm \lambda}(t))&\nonumber\\
 &=\sum\limits_{n \in {\cal N}} {\sum\limits_{b \in {\cal O}(n)} {\sum\limits_{f \in {\cal F}} {\sum\limits_{s \in {\cal N}_s^f} {\sum\limits_{d \in {\cal N}_d^f} {\tilde x_{nb}^{fsd}(t)\left( {Q_n^{fsd}(t) - Q_b^{fsd}(t)} \right)} } } } }&\nonumber \\
 &- \sum\limits_{n \in {\cal N}} {\sum\limits_{b \in {\cal O}(n)} [ } \sum\limits_{f \in {\cal F}} {A_b(t)\tilde P_b^Rx_{nb}^f\left( t \right)}
 + A_n(t)e^{\hat{p}_{nb}^T\left( t \right)}]&
\end{flalign}


Let ${\cal D}({\bm \lambda}(t),{\bm \rho}(t))$ be the solution of  \textbf{P4}, the dual problem of \textbf{P4} is
\begin{eqnarray}\label{opt3}
\mathop {\mbox{max}}_{{\bm \lambda}(t)\succeq 0,{\bm \rho}(t)\succeq 0} && {\cal D}({\bm \lambda}(t),{\bm \rho}(t))
\end{eqnarray}
and can be solved by a gradient projection method, i.e.,
\begin{equation}\label{lamba update}
 {\lambda _n}\left( {{t_{i + 1}}} \right) = {\left[ {{\lambda _n}\left( {{t_i}} \right) + {\kappa _\lambda }\left( {{t_i}} \right)\frac{{\partial {\cal L}\left( {\bm \chi \left( {{t_i}} \right),\bm \lambda \left( {{t_i}} \right),\bm \rho \left( {{t_i}} \right)} \right)}}{{\partial {\lambda _n}\left( {{t_i}} \right)}}} \right]^ + }
\end{equation}
\begin{equation}\label{rho update}
\rho _m^f\left( {{t_{i + 1}}} \right) = {\left[ {\rho _m^f\left( {{t_i}} \right) + {\kappa _\rho }\left( {{t_i}} \right)\frac{{\partial {\cal L}\left( {\bm \chi \left( {{t_i}} \right),\bm \lambda \left( {{t_i}} \right),\bm \rho \left( {{t_i}} \right)} \right)}}{{\partial \rho _m^f\left( {{t_i}} \right)}}} \right]^ + }
\end{equation}
where
\begin{flalign}
&\frac{{\partial {\cal L}\left( {\bm \chi \left( {{t_i}} \right),\bm \lambda \left( {{t_i}} \right),\bm \rho \left( {{t_i}} \right)} \right)}}{{\partial {\lambda _n}\left( {{t_i}} \right)}}&\nonumber\\
& ={g_n}\left( t \right) - {d_n}\left( t \right) + \sum\limits_{f \in {\cal F}} {{1_{n \in {\cal N}_s^f}}\tilde P_f^Sr_n^f(t)}& \nonumber\\
&+ \sum\limits_{b \in {\cal O}(n)} e^{\hat{p}_{nb}^T(t)}  + \sum\limits_{a \in {\cal I}(n)} {\sum\limits_{f \in {\cal F}} {x_{an}^f(t)} }  - {y_n}\left( t \right)&
\end{flalign}
\begin{flalign}
&\frac{{\partial {\cal L}\left( {\bm \chi \left( {{t_i}} \right),\bm \lambda \left( {{t_i}} \right),\bm \rho \left( {{t_i}} \right)} \right)}}{{\partial \rho _m^f\left( {{t_i}} \right)}}&\notag\\
&= H\left( {\underline{{\cal N}}_{sm}^f|{\cal N}_s^f - _{sm}^f} \right){\rm{ - }}\log \left( {{{\left( {2\pi e} \right)}^{|\underline{{\cal N}}_{sm}^f|}}\prod\limits_{n \in \underline{{\cal N}}_{sm}^f} {D_n^f\left( t \right)} } \right)&\notag\\
&{\rm{ - }}\sum\limits_{n \in \underline{{\cal N}}_{sm}^f} {r_n^f\left( t \right)}&
\end{flalign}
where $t_i$ represents the iteration number at time slot t, ${\kappa _\lambda }\left( {{t_i}} \right)$ and ${\kappa _\rho }\left( {{t_i}} \right)$ are the step sizes.

\subsection{Solving \textbf{P4}}
Our problem \textbf{P4} can be divided into five independent subproblems, in the following subsections, we will give the full subscriptions.

(1) \textbf{Energy harvesting on EH node}
For each EH node $n \in {{\cal N}_H}$, combining the first term of the RHS of \eqref{L1} with the constraint \eqref{harvesting_decision} and \eqref{limitedcapacityeh}, we have the optimization problem of ${e_n}(t)$ as follows:
\begin{eqnarray}\label{Energy Harvesting}
\mathop{\mbox{minimize}}_{e_n(t)}&&\left( {{{E_n^H}}(t) - \theta _{n}^{{e_H}}} \right) {e_n}(t)\\
\mbox{subject to}&&0\le {e_n}(t) \le {h_n}(t)\nonumber\\
&&{{E_n^H}}(t)+{{e_n}\left( t \right)}  \le{ \theta _{n}^{{eH}}}\label{SolveEnergy Harvesting}
\end{eqnarray}
\textbf{Remark 3.2}
\eqref{SolveEnergy Harvesting} indicates that all the incoming energy is stored if there is enough room in the energy buffer according to the limitation imposed by ${\theta _n^{e_H}}$, and otherwise it stores all the energy that it can, filling up the battery size of ${\theta _n^{e_H}}$. Hence, ${{E_n^H}}(t) <{\theta _n^{{e_H}}}$ for all $t$,
which means that
our algorithm can be implemented with finite energy storage capacity ${\theta _n^{{e_H}}}$ at node $n \in {\cal N}_H$.

(2) \textbf{Energy harvesting and Battery charging on ME nodes}
Combining the ${\cal L}_1$ with the constraints \eqref{harvesting_decision} and \eqref{charging_rate constraint}, for each ME node $n \in {{\cal N}_M}$, we have the optimization problem of ${e_n}(t)$ and ${g_n}(t)$ as follows:
\begin{eqnarray}\label{EPBCD}
\mathop{\mbox{minimize}}_{e_n(t),g_n(t)}&&\left({ E_n^M\left( t \right) - \theta _n^M+{\lambda _n}} \right){g_n}\left( t \right)\notag\\
 &&+\left({ E_n^M\left( t \right) - \theta _n^M} \right){e_n}\left( t \right)\\
\mbox{subject to}&&0\le {e_n}(t) \le {h_n}(t)\nonumber\\
&& 0\le {g_n}\left( t \right) \le g_n^{\max }\nonumber
\end{eqnarray}
(3) \textbf{Energy purchase and Battery discharging on ME nodes}
Combining the ${\cal L}_2$ with the constraints \eqref{discharging_rate constraint} and \eqref{energy supplied constraint}, for each ME node $n \in {{\cal N}_M}$, we have the optimization problem of ${d_n}(t)$ and ${y_n}(t)$ as follows:
\begin{eqnarray}\label{EPBCD1}
\mathop{\mbox{minimize}}_{{d_n}(t),{y_n}(t)}&&\left({V(1-\varpi_1)\varpi_2{P_n^{G}}(t)-{\lambda _n}}\right){y_n}(t)\notag\\
&&-\left({ E_n^M\left( t \right) - \theta _n^M+{\lambda _n}} \right){d_n}\left( t \right)\\
\mbox{subject to}&& 0\le {d_n}\left( t \right) \le d_n^{\max }\nonumber\\
&& 0\le {y_n}\left( t \right) \le y_n^{\max }\nonumber
\end{eqnarray}
(4) \textbf{Energy purchase and Battery charge/discharge on EG node}
Combining the ${\cal L}_3$ with the constraints \eqref{charging_rate constraint}-\eqref{energy supplied constraint} for each node $n \in {{\cal N}_G} $, we get the optimization problem of ${y_n}(t)$ , ${g_n}(t)$ and ${d_n}(t)$ as follows:
\begin{eqnarray}\label{EPBCD2}
\mathop{\mbox{minimize}}_{{y_n}(t),{d_n}(t),{g_n}(t)}&&\left({ E_n^G\left( t \right) - \theta _n^G+{\lambda _n}} \right)({g_n}\left( t \right)-{d_n}\left( t \right))\nonumber\\
&&+\left({V(1-\varpi_1)\varpi_2{P_n^{G}}(t)-{\lambda _n}}\right){y_n}(t)\\
\mbox{subject to}&& 0\le {g_n}\left( t \right) \le g_n^{\max }\nonumber\\
&& 0\le {d_n}\left( t \right) \le d_n^{\max }\nonumber\\
&& 0\le {y_n}\left( t \right) \le y_n^{\max }\nonumber
\end{eqnarray}
(5) \textbf{Distortion optimization}
Combining the ${\cal L}_4$ with the constraints \eqref{distortion_constraint}, for each session $f \in {{\cal F}}$, at node $n \in {{\cal N}_s^f}$, we get the optimization problem of ${D_n^f}(t)$ as follows:
\begin{eqnarray}\label{D}
\mathop{\mbox{maximize}}_{{D_n^f}(t)}&& V\varpi_1U_n^f\left( {D_n^f(t)} \right)+ \log \left( {D_n^f(t)} \right)\sum\limits_{m:n \in \underline{{\cal N}}_{sm}^f} {\rho _m^f\left( t \right)}\nonumber\\
\mbox{subject to}&& D_{\min} \le {D_n^f}\left( t \right) \le D_{\max }
\end{eqnarray}
(6) \textbf{Source rate control}
Combining the ${\cal L}_5$ with the constraints \eqref{source_rate_constraint}, for each session $f \in {{\cal F}}$, at node $n \in {{\cal N}_s^f}$, we get the optimization problem of ${r_n^f}(t)$ as follows:
\begin{eqnarray}\label{SR}
\mathop{\mbox{maximize}}_{{r_n^f}(t)}&&[ \sum\limits_{m:n \in \underline{{\cal N}}_{sm}^f} {\rho _m^f\left( t \right)}- \sum\limits_{d \in {\cal N}_d^f} {Q_n^{fnd}(t)} ]  r_n^f(t)\nonumber\\
&& + A_n(t)\tilde P_f^S r_n^f(t)\notag\\
\mbox{subject to}&& 0 \le {r_n^f}\left( t \right) \le R_{\max }
\end{eqnarray}
(7) \textbf{Information rate, Physical rate and Power allocation}
We have the optimization problem of ${\tilde{\bm x}}(t)$, ${{\bm x}}(t)$ and ${\bm {\hat{p}}}^{T}(t)$ as follows:
\begin{flalign}\label{DRNL}
\mathop{\mbox{maximize}}_{{\tilde{\bm x}}(t),{{\bm x}}(t),{\bm {\hat{p}}}^{T}(t)}&\;\;{\cal L}_6({\bm x}(t),\tilde{{\bm x}}(t),{\bm {\hat{p}}}^{T}(t),{\bm \lambda}(t))  \\
\mbox{subject to}&\;\;\tilde x_{nb}^{fsd}(t) \le x_{nb}^f\left( t \right),\;\forall n \in {\cal N},b \in {\cal O}(n),\label{constraint1}\\
&\;\;\;\;\;\;\;\;\;\;\quad f \in {\cal F},s \in {\cal N}_s^f,d \in {\cal N}_d^f\nonumber\\
&\;\;\sum\limits_{b \in {\cal O}\left( n \right)} {\tilde{x}  _{{nb}}^{fsd}\left( t \right)} \leq {{{Q}_n^{fsd}}}\left( t \right),\nonumber\\
&\;\;\;\;\;\;\;\;\;\;\forall n \in {\cal N},f \in {\cal F},s \in {\cal N}_s^f,d \in {\cal N}_d^f\nonumber\\
&\;\;\sum\limits_{f \in {\cal F}} {x_{nb}^f\left( t \right)}  \le {\tilde{C}_{nb}}\left( t \right),\;\forall n \in {\cal N},b \in {\cal O}(n)\notag\\
&\;\; \sum\limits_{b \in {\cal O}\left( n \right)} e^{{\hat{p}_{nb}^T}(t)}  \le {P_n^{\max }},\;\; \forall n \in {{\cal N}}\nonumber
\end{flalign}
Next, we replace the variable $\tilde{x}_{nb}^{fsd}(t)$ with the variable $x_{nb}^f(t)$ due to the constraint \eqref{constraint1}, we have
\begin{eqnarray}
 &&{\cal L}_6({\bm x}(t),{\bm {\hat{p}}}^{T}(t),{\bm \lambda}(t))\nonumber\\
 &&=\sum\limits_{n \in {\cal N}} {\sum\limits_{b \in {\cal O}(n)} }[  \sum\limits_{f \in {\cal F}} {w_{nb}^f(t)x_{nb}^f\left( t \right)}
 + A_n(t)e^{\hat{p}_{nb}^T\left( t \right)}]
\end{eqnarray}
where
\begin{equation}\label{w}
  w_{nb}^f\left( t \right) = \sum\limits_{s \in {\cal N}_s^f} {\sum\limits_{d \in {\cal N}_d^f} {(Q_n^{fsd}(t) - Q_b^{fsd}(t))} }  + {A_b}\left( t \right)\tilde P_b^R
\end{equation}
We define ${ W_{nb}^f}(t) \buildrel \Delta \over = {\left[{w}_{nb}^f(t) - \sum\limits_{s \in {\cal N}_s^f} {\sum\limits_{d \in {\cal N}_d^f} \epsilon } \right]^ + }$,\\
where $\epsilon=l_{\max}{X}_{\max}+R_{\max}$.
Note that
the \emph{data-availability} constraint \eqref{Data_availability}  is always satisfied by introducing $\epsilon$, we will show it in Part D  of Theorem 2. Thus, we can get rid of this constraint.

\textbf{Transmission Power Allocation Component}
For each node $n$, find any
${f^*} \in \arg {\max _f}\; W_{nb}^f(t)$.
Define $ W_{nb}^{*}(t) = {{{\max }_f}\; W_{nb}^f(t)}$ as the corresponding optimal weight of link $\left( {n,b} \right)$.
Observe the current channel state ${\bm S}(t)$, and select the transmission powers
${\bm \hat{p}}^{T*}$
by solving the following optimization problem :
\begin{eqnarray}\label{opeq}
\mathop{\mbox{maximize}}_{{\bm {\hat{p}}}^{T}(t)}&&\sum\limits_{n \in {\cal N}} {\sum\limits_{b \in {\cal O}\left( n \right)} {\left( { W_{nb}^*(t){\tilde{C}_{nb}(t)} + {A_{n}}(t)e^{{\hat{p}_{nb}}(t)}} \right)} }  \nonumber\\
\mbox{subject to}&&\sum\limits_{b \in {\cal O}\left( n \right)} e^{{\hat{p}_{nb}^T(t)}}  \le {P_n^{\max }},\forall n \in {\cal N}
\end{eqnarray}

\textbf{Session Scheduling}
The data of session ${f^*}$  is
selected for routing over link $(n, b)$ whenever
$ W_{nb}^{{f^*}}(t) > 0$.
That is, if $ W_{nb}^{{f^*}}(t) > 0$, set $x _{nb}^{{f^*}}(t)={\tilde{C}_{nb}}\left({\bm \hat{p}}^{T*}, {\bm S}(t) \right)$.
Next, for each session $f \in \cal F$, if $f=f^*$ and $Q_n^{fsd}(t) - Q_b^{fsd}(t)+ {A_b}\left( t \right)\tilde P_b^R - \epsilon>0$,let
$\tilde{x}_{nb}^{fsd}={\tilde{C}_{nb}}\left({\bm \hat{p}}^{T*}, {\bm S}(t) \right)$.

\textbf{Remark 3.3}:
To  distributively solve the problem \eqref{opeq}, we propose a distributed iterative algorithm based
on block coordinate descent (BCD) method.

\section{Performance analysis}

Now, we analyze the performance of our proposed algorithm.
To start with, we assume that there exists $\delta>0$  such that
\begin{equation}\label{PowerFeature1}
{C_{nb}}\left( {{{\bm p}^T}\left( t \right),{\bm S}\left( t \right)} \right) \le \delta BW{p_{nb}^T}\left( t \right), \forall n \in {\cal N}, \forall b \in {\cal O}\left( n \right).
\end{equation}
We define $\tilde{p}_S=\mathop {\min }\limits_f \tilde p_f^S$, $\tilde{p}_R=\mathop {\min }\limits_n \tilde p_n^R$, $\sigma=\min\{\tilde p_S, \tilde p_R, 1\}$, 
and \[\beta=\varpi_1
{\sup _{{D_{\min }} \le D_n^f\left( t \right) \le {D_{\max }}}}\frac{{U_n^f\left( {D_n^f\left( t \right)} \right) - U_n^f\left( {{D_{\max }}} \right)}}{{\log \left( {{D_{\max }}} \right) - \log \left( {D_n^f\left( t \right)} \right)}}\].

\textbf{Theorem 2}:
The perturbed variables are chosen as follows:
\begin{eqnarray}
  \theta _n^{eH} &=& {\max\{\frac{{1}}{\sigma},\delta BW\}} N_sN_d \beta V+ P_{n,\max }^{Total}, n\in {\cal N}_H, \label{thetaeh} \\
  \theta _{n}^{eG} &=& \frac{{1}}{{\sigma}}\beta V+ d_n^{\max}, n\in {\cal N}_G,\label{thetaeg}\\
  \theta _{n}^{eM} &=& \frac{{1}}{{\sigma}}\beta V+ d_n^{\max}, n\in {\cal N}_M,\label{thetaem}
\end{eqnarray}

Then, implementing the algorithm  with any fixed
parameter $V>0$ for all time slots, we have the following
performance guarantees:

(\textbf{A}).
Suppose the initial data queues and the initial energy queues satisfy:
\begin{eqnarray}
Q_n^{fsd}\left( 0\right) &\le& \beta V+ R_{\max },\forall n,f,s \in {\cal N}_s^f,d \in {\cal N}_d^f\label{Qbound0} \\
E_{n}^H\left( 0\right) &\le& \theta _{n}^{eH},\quad n \in {{\cal N}_H}\label{ehu0}\\
E_{n}^G\left( 0\right) &\le& \theta _{n}^{eG}+g_n^{\max},\quad n \in {{\cal N}_G}\label{egu0}\\
E_n^M\left( 0\right) &\le&  \theta _{n}^{eM}+g_n^{\max}+h_{\max},\quad n \in {{\cal N}_M}\label{emu0}
\end{eqnarray}
then, the data queues and the energy queues of all nodes for all time slots $t$ are always
bounded as
\begin{eqnarray}
Q_n^{fsd}\left( t \right) &\le& \beta V+ R_{\max },\forall n,f,s \in {\cal N}_s^f,d \in {\cal N}_d^f\label{Qbound} \\
E_{n}^H\left( t \right) &\le& \theta _{n}^{eH},\quad n \in {{\cal N}_H}\label{ehu}\\
E_{n}^G\left( t \right) &\le& \theta _{n}^{eG}+g_n^{\max},\quad n \in {{\cal N}_G}\label{egu}\\
E_n^M\left( t \right) &\le& \theta _{n}^{eM}+g_n^{\max}+h_{\max},\quad n \in {{\cal N}_M}\label{emu}
\end{eqnarray}

(\textbf{B}).  The objective function value of the problem \textbf{P1} achieved by the proposed algorithm  satisfies the bound
\begin{equation}\label{objective_value}
\overline{O} \ge {O^*} - \frac{{\tilde{B}}}{V}
\end{equation}
where ${O^*}$ is the optimal value of  the problem \textbf{P1},
and $\tilde{B}=B+NFN_s N_d\epsilon  l_{\max}X_{\max}$.

(\textbf{C}).  When node $n \in {{\cal N}_H}$ allocates nonzero power for data sensing, data transmission and/or data reception, we have:
\begin{equation}\label{ea1}
E_{n}^H\left( t \right) \ge P_{n,\max }^{Total}, n \in {{\cal N}_H}.
\end{equation}

When the battery on node $n \in {{\cal N}_G}\bigcup{{\cal N}_M}$ is discharging, we have:
\begin{equation}\label{ea2}
 E_{n}^G\left( t \right) \ge d_n^{\max }, n \in {{\cal N}_G},
 \end{equation}
 \begin{equation}\label{ea3}
 E_{n}^M\left( t \right) \ge d_n^{\max }, n \in {{\cal N}_M}.
\end{equation}

(\textbf{D}). For each node $n \in {{\cal N}}$, when the node $n$ transmits the $f$-th session from source $s \in {\cal N}_s^f$ to sink $d \in {\cal N}_d^f$,we have:
 \begin{equation}\label{ea4}
Q_{n}^{fsd}\left( t \right) \ge l_{\max}{X}_{\max}.
\end{equation}

\textbf{Proof}: Please see Appendix A-D.

\textbf{Remark 4.1}:
\textbf{Theorem 2} shows that a control parameter $V$  enables an explicit trade-off between the average objective value and queue backlog. Specifically, for any $V >0$,
the proposed distributed algorithm CLEAR can achieve a time average objective that is within ${\cal {O}}(1/V)$ of the optimal objective, while ensuring that the average data/energy queues have upper bounds of ${\cal  {O}}(V)$.
In the next section, the simulations will verify the theoretic claims.

\textbf{Remark 4.2}: The inequations \eqref{ea1}-\eqref{ea4}  guarantees that the  \emph{energy-availability} constraints \eqref{EMenergy_availability},  \eqref{EHenergy_availability} , \eqref{EGenergy_availability}, and the \emph{data-availability} constraints \eqref{Data_availability} are satisfied for all nodes and all times.

\section{Simulation Results}\label{Simulation}

\begin{figure}
\centering
\includegraphics[width=1.5in]{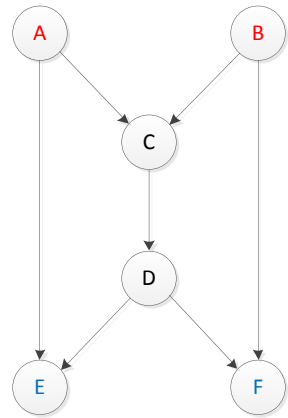}
\centering
\caption{Network topology.}
\centering
\label{fig:Topology}
\centering
\end{figure}

In this section, we provide some numerical examples to verify the performance of our alforithm. We consider the network
topology in Fig.\ref{fig:Topology}, which has 6 nodes, 7 links. Moreover, we assume there is only one multicast session, the nodes $\{A,B\}$ are the two correlated sources and the nodes $\{E,F\}$ are the sinks.
Throughout,
the form of the rate utility function is set as $U_n^f(D_n^f(t))=\log(1-D_n^f(t))$.
The form of the electricity cost function is set as $P_n^{G}(t)=S_n^{G}(t)$.

\subsection{Default simulation setting}
Set $\mathcal{N}_H=\{A,D\}, \mathcal{N}_G=\{B\}, \mathcal{N}_M=\{C\}$ as the default node scenario.
Set several default values as in TABLE II:
\begin{table}[!hbp]
   \caption{\label{tab:table2}Values of parameters}
   \centering
\begin{tabular}{lll}
  \toprule
  Parameter & Value\\
  \midrule
  $R_{\max}$ & 10\\
  $D_{\min}$ & 0.01\\
  $D_{\max}$ & 0.8\\
  $P_n^{\max},\forall n \in {\cal N}$ & 8\\
  $N_{nb}, \forall (n,b) \in {\cal L}$ & $5\times 10^{-13}$\\
  $BW$ & 10\\
  $X_{\max}$ & 10\\
  ${\tilde P}_f^S, \forall f \in {\cal F}$ & 0.1\\
  ${\tilde P}_n^R, \forall n \in {\cal N}$ & 0.05\\
  $g_n^{\max }, \forall n \in {{{\cal N}_G} \cup {{\cal N}_M}}$ & 15\\
  $d_n^{\max }, \forall n \in {{{\cal N}_G} \cup {{\cal N}_M}}$ & 15\\
  $y_n^{\max }, \forall n \in {{{\cal N}_G} \cup {{\cal N}_M}}$ & 25\\
  $l_{\max}$ & 2\\
  $\varpi_1$ & 0.7\\
  $\varpi_2$ & 0.1\\
  $\beta$ & 2.8\\
 $\sigma$ & 0.05\\
  $\delta$ & 2\\
    \bottomrule
\end{tabular}
\end{table}

Therein, we assume the $S_{nb}^C(t)=d_{nb}^{-4}$. The energy-harvesting vector ${\bm S}^H(t)$  has
independent entries that are uniformly distributed in $[0, 10]$ at ME nodes and $[0, 50]$ at EH nodes.
The electricity price vector ${\bm S}^{G}(t)$ has
independent entries that are uniformly distributed in $[S_{\min}^{G}, S_{\max}^{G}]$
with $S_{\min}^{G}=0.5$, $S_{\max}^{G} =1$ as default values.

The initial queue sizes and the queue upper bounds are set according to Theorem 2. Therein, we set all the initial queue sizes to be zero.
According to \eqref{thetaeh}-\eqref{thetaem} in Theorem 2, we set
$\theta _n^{{eH}} = 224V + 38$,$\theta _n^{{eG}} = 56V + 15$,$\theta _n^{{eM}} = 56V + 15$. According to \eqref{Qbound}-\eqref{emu}, for $\forall n,f,s \in {\cal N}_s^f,d \in {\cal N}_d^f$, each data queue $Q_n^{fsd}$ has an upper bound $2.8V + 10$,
and the battery buffer sizes of each EH node, of each EG node, and of each ME node are set to be $224V + 38$, $56V + 30$ and $56V + 40$, respectively.
\begin{figure}[t]
\centering
\includegraphics[width=3.5in]{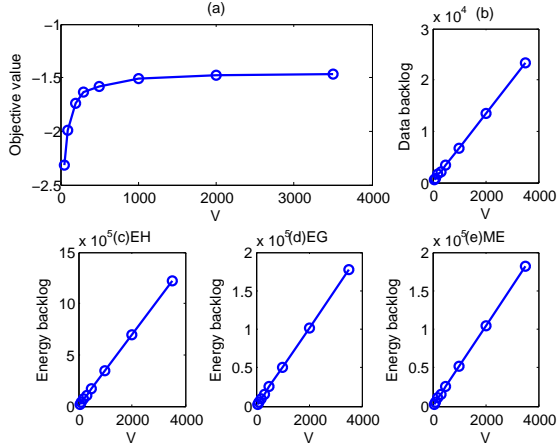}
\caption{Verification of Theorem 1.}
\label{fig_Verifying_theoretic_claims}
\end{figure}
\subsection{Algorithm performance evaluation}
We simulate $V =[50,100,200,300,500,1000,2000,3500]$.
In all simulations, the simulation time is $7\times 10^{4}$ time slots.

We first examine the effect of parameter V we described in Theorem 2, which can achieve a tradeoff $(V,1/V)$ between queue sizes and the gap from the optimal value.
In Fig. \ref{fig_Verifying_theoretic_claims},(a) describes the gap from the optimal value, and we can see that as $V$ increases, the time average optimization objective value keep increasing and converge to very close to  the optimum. This confirms the results of \eqref{objective_value}.
(b) describes the data queue sizes, we see that as $V$ increases, the queue size
keeps increasing. (c)-(e)describes the energy queue sizes, we observe that the queue size increases as $V$ increases. These confirm the results of \eqref{Qbound}-\eqref{emu}.

\begin{figure}[t]
\centering
\includegraphics[width=3.5in]{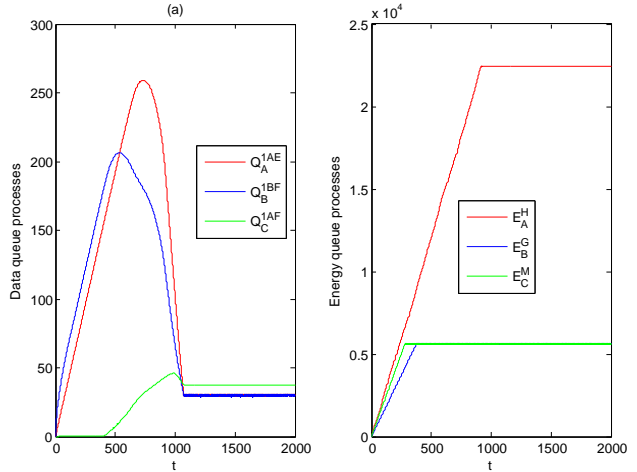}
\caption{Detailed verification of the queueing bounds.}
\label{fig_detailed_verification_queue_bound}
\end{figure}
Fig \ref{fig_detailed_verification_queue_bound} shows three data queue processes and three energy queue processes under $V=100$.
(a) shows the data queue
processes of node $A$ from source node $A$ to sink node $E$, node $B$ from source node $B$ to sink node $F$ and node $C$ from source node $A$ to sink node $F$, respectively. (b) shows the energy queue
processes for EH node $A$, EG node $B$ and ME node $C$, respectively.
It can be seen that the queue sizes are all bounded with upper bound given in Theorem 2.






\section{Conclusions}
It is a huge challenge to
remain the perpetual operation for WMSN, due to the extensive energy consumption.
In this paper, we consider the coexistence of renewable
energy and grid
power supply for WMSN.
We consider multiple energy consumptions, and heterogeneous energy supplies in the system model, and formulate a discrete-time stochastic cross-layer optimization problem for the multiple multicast with DSC and NC to achieve the optimal reconstruct distortion at sinks  and the minimal cost of purchasing electricity from electricity grid.
Based on the Lyapunov drift-plus-penalty with perturbation technique and dual decomposition technique, we propose a fully distributed and
low-complexity cross-layer algorithm only requiring knowledge of the instantaneous system state.
The theoretic proof and the simulation results show that a parameter $V$ enables an explicit trade-off between the optimization objective and queue backlog.
In the future, we are interested in delay
reduction by modifying the queueing disciplines.

\appendices
\section{Proof of Part (\textbf A) in Theorem 2}
For $t=0$, we can easily have \eqref{Qbound}, then we assume \eqref{Qbound} is hold at time slot $t$, next we will show that it holds at $t+1$.

\textbf{Case 1}: If queue $Q_n^{fsd}(t)$ doesn't receive any data at time $t$, we have
$Q_n^{fsd}\left( {t + 1} \right) \le Q_n^{fsd}\left( {t } \right) \le  {\beta}V + R_{\max }$.

\textbf{Case 2}: If queue $Q_n^{fsd}(t)$ receives the endogenous data from other nodes
$a \in {\cal I}\left( n \right)$, we can get from the session scheduling component  that
$Q_a^{fsd}(t) - Q_n^{fsd}(t)+ {A_n}\left( t \right)\tilde P_n^R - \epsilon \ge 0$.
By plugging the definition of $\epsilon$, and
due to $A_{n}(t)\leq 0$ and $\tilde P_n^R >0$, we have
  \begin{equation}\label{pf2}
    Q_n^{fsd}\left( t \right) \le {{Q}}_a^{fsd}\left( t \right) -\left( { R_{\max } + {l_{\max }}{X_{\max }}} \right ).
  \end{equation}
Plugging \eqref{Qbound} into \eqref{pf2}, we have
\begin{eqnarray}\label{add1}
  Q_n^{fsd}\left( t \right)&\le&  \beta V+R_{\max }- \left( { R_{\max } + {l_{\max }}{X_{\max }}}\right )\nonumber\\
  &=&\beta V-{l_{\max }}{X_{\max }}.
\end{eqnarray}
At every slot, the node can receive the amount of session $f$ data from source $s \in {\cal N}_s^f$ to sink $d \in {\cal N}_d^f$ at most $ R_{\max } + {l_{\max }}{X_{\max }} $. So
\begin{eqnarray} \label{add2}
  Q_n^{fsd}\left( {t + 1} \right) &\le& Q_n^{fsd}\left( t \right) + {l_{\max }}{X_{\max }} + R_{\max }.
\end{eqnarray}
Combing \eqref{add1} and \eqref{add2}, we have
$Q_n^{fsd}\left( {t + 1} \right) \le \beta V + r_f^{\max }$.

\textbf{Case 3}:
\emph{Lemma} 1: For any vector $\bm \lambda^{*}$,$\bm \rho^{*}$ that maximizing \eqref{opt3}, we have
\begin{equation}\label{rhob}
 \sum\limits_{m:n \in \underline{{\cal N}}_{sm}^{f}} {\rho _m^{f*}\left( t \right)}\le \beta V
\end{equation}
\begin{equation}\label{lambdab}
 \lambda_n^{*} \le \frac{{1}}{{\sigma}}\beta V
\end{equation}
\textbf{Proof}: please see Appendix E

If queue $Q_n^{fsd}(t)$ only receives the new local data, in other words, the node $n=s$, and exits an optimal solution $\bm r^{*}>0$ of \eqref{SR}
, we have the following:
\[ \sum\limits_{m:n \in \underline{{\cal N}}_{sm}^f} {\rho _m^{f*}\left( t \right)}- \sum\limits_{d \in {\cal N}_d^f} {Q_n^{fnd}(t)}  + A_n(t)\tilde P_f^S \ge 0\]
From \eqref{rhob}, we get
\begin{eqnarray}
  \sum\limits_{d \in {\cal N}_d^f} {Q_n^{fnd}(t)} &\le& \sum\limits_{m:n \in \underline{{\cal N}}_{sm}^f} {\rho _m^{f*}\left( t \right)}+ A_n(t)\tilde P_f^S\nonumber\\
   &\le& \beta V
\end{eqnarray}
i.e.,$Q_n^{fsd}(t) \le \beta V$. Hence $Q_n^{fsd}(t+1) \le Q_n^{fsd}(t)+R_{\max} \le \beta V+R_{\max}$

To sum up the above, we complete the proof of \eqref{Qbound}.

From \textbf{Remark 3.2} we can have \eqref{ehu}.

From \eqref{EPBCD1}, we obtain that if $E_n^G\left( t \right) \le \theta _n^G-{\lambda _n}$, i.e., $E_n^G\left( t \right) \le \theta _n^G$, the EG node is charging, so $E_n^G\left( t+1 \right) \le E_n^G\left( t \right)+g_n^{\max} \le \theta _n^G+g_n^{\max}$.

From \eqref{EPBCD}, we obtain that if $E_n^M\left( t \right) \le \theta _n^M-{\lambda _n}$, i.e., $E_n^M\left( t \right) \le \theta _n^M$, the ME node is charging, so $E_n^M\left( t+1 \right) \le E_n^M\left( t \right)+g_n^{\max}+h_{\max} \le \theta _n^G+g_n^{\max}+h_{\max}$.

So, this completes the proof of \textbf{Theorem 2}(\textbf{A}). $\square$
\section{Proof of Part (\textbf B) in Theorem 2}

Before giving the proof of  Part (B) in Theorem 2, we give \emph{Lemma} 2 and its proof.
At time $t$, we denote $\alpha_{n,fsd}^Q\left( t \right)$, $\alpha_n^{eH}(t)$, $\alpha_n^{eG}(t)$, $\alpha_n^{eM}(t)$ and $
\mu_{n,fsd}^Q\left( t \right)$, $\mu_n^{eH}(t)$, $\mu_n^{eG}(t)$, $\mu_n^{eM}(t)$as the input and output of the queue $Q_n^{fsd}\left( t \right)$, $E_n^H(t)$, $E_n^G(t)$ and $E_n^M(t)$ for $\forall f \in {\cal F},s \in {\cal N}_s^f, d \in {\cal N}_d^f,n \in {\cal N},b \in {\cal O}\left( n \right)$, respectively.

\emph{Lemma} 2: We assume  ${\phi ^*}$ is the optimal solution to the following problem \eqref{phi} and the ${\bm Z}(t)$ takes non-negative values from a
finite but arbitrarily large set $\cal Z$,
\begin{figure*}
\begin{eqnarray}\label{phi}
&&\mbox{maximize}\quad V\sum\limits_{{z_i}\in \cal Z} {{\pi _{{z_i}}}\sum\limits_{k \in {\cal K}} {a_k^{\left( {{z_i}} \right)}} O} \left( {{z_i},{\bm \chi } _k^{\left( {{z_i}} \right)}} \right)\\
&&\mbox{subject to}\notag\\
&&\sum\limits_{{z_i}\in \cal Z} {{\pi _{{z_i}}}} \sum\limits_{k \in {\cal K}} {a_k^{\left( {{z_i}} \right)}\alpha_{n,fsd}^Q} \left( {{z_i},{\bm \chi } _k^{\left( {{z_i}} \right)}} \right) = \sum\limits_{{z_i}\in \cal Z} {{\pi _{{z_i}}}} \sum\limits_{k \in {\cal K}} {a_k^{\left( {{z_i}} \right)}\mu_{n,fsd}^Q} \left( {{z_i},{\bm \chi } _k^{\left( {{z_i}} \right)}} \right),\nonumber\\
&&\sum\limits_{{z_i}\in \cal Z} {{\pi _{{z_i}}}} \sum\limits_{k \in {\cal K}} {a_k^{\left( {{z_i}} \right)}\alpha_n^{eH}} \left( {{z_i},{\bm \chi } _k^{\left( {{z_i}} \right)}} \right) = \sum\limits_{{z_i}\in \cal Z} {{\pi _{{z_i}}}} \sum\limits_{k \in {\cal K}} {a_k^{\left( {{z_i}} \right)}\mu_n^{eH}} \left( {{z_i},{\bm \chi } _k^{\left( {{z_i}} \right)}} \right),\notag\\
&&\sum\limits_{{z_i}\in \cal Z} {{\pi _{{z_i}}}} \sum\limits_{k \in {\cal K}} {a_k^{\left( {{z_i}} \right)}\alpha_n^{eG}} \left( {{z_i},{\bm \chi } _k^{\left( {{z_i}} \right)}} \right)= \sum\limits_{{z_i}\in \cal Z} {{\pi _{{z_i}}}} \sum\limits_{k \in {\cal K}} {a_k^{\left( {{z_i}} \right)}\mu_n^{eG}} \left( {{z_i},{\bm \chi } _k^{\left( {{z_i}} \right)}} \right),\notag\\
&&\sum\limits_{{z_i}\in \cal Z} {{\pi _{{z_i}}}} \sum\limits_{k \in {\cal K}} {a_k^{\left( {{z_i}} \right)}\alpha_n^{eM}} \left( {{z_i},{\bm \chi } _k^{\left( {{z_i}} \right)}} \right)= \sum\limits_{{z_i}\in \cal Z} {{\pi _{{z_i}}}} \sum\limits_{k \in {\cal K}} {a_k^{\left( {{z_i}} \right)}\mu_n^{eM}} \left( {{z_i},{\bm \chi } _k^{\left( {{z_i}} \right)}} \right),\notag\\
&&\sum\limits_{n \in {\cal \underline{N}}_s^f} {r_{n,k}^{f(z_i)}} \ge H\left( {{\cal \underline{N}}_s^f|{\cal N}_s^f-{\cal \underline{N}}_s^f} \right)-\log((2{\pi}e)^{|{\cal \underline{N}}_s^f|}\prod\limits_{n \in {\cal \underline{N}}_s^f} {D_{n,k}^{f({z_i})}}),\forall {\cal \underline{N}}_s^f \subseteq {\cal N}_s^f\notag\\
&&0 \le r_{n,k}^{f({z_i})} \le R_{\max },D_{\min} \le D_{n,k}^{f({z_i})} \le D_{\max },\forall n \in {\cal N}_s^f\notag\\
&& \tilde{x}_{nb,k}^{fsd({z_i})} \le x_{nb,k}^{f({z_i})}, \sum\limits_{f \in {\cal F}} {x_{nb,k}^{f({z_i})}}  \le C_{{nb,k}}^{({z_i})}, \sum\limits_{b \in {\cal O}\left( n \right)} {p_{nb,k}^{T({z_i})}}  \le P_n^{\max }, \forall b \in {\cal O}\left( n \right)\notag\\ 
&& 0 \le g_{n,k}^{({z_i})} \le g_n^{\max },0 \le d_{n,k}^{({z_i})} \le d_n^{\max }, 0 \le y_{n,k}^{({z_i})} \le y_n^{\max },\forall n \in {{{\cal N}_G \cup {\cal N}_M}},\notag\\
&&y_{n,k}^{({z_i})}=g_{n,k}^{({z_i})}-d_{n,k}^{({z_i})}+ p_{n,k}^{Total(z_i)},\forall n \in {{{\cal N}_G \cup {\cal N}_M}},\notag\\
&&0 \le e_{n,k}^{({z_i})} \le h_{n,k}^{({z_i})},\forall n \in {{{\cal N}_H \cup {\cal N}_M}},\notag\\
&& \sum\limits_{k \in {\cal K}} {a_k^{\left( {{z_i}} \right)}}  = 1, \;\;\;\;a_k^{\left( {{z_i}} \right)} \ge 0,\;\;\;\;{\bm \chi } _k^{\left( {{z_i}} \right)} \in {{\bm \chi } ^{{z_i}}},\notag\\
&& \forall k, {z_i},n,f,s\in {\cal N}_s^f,d\in {\cal N}_d^f,\notag
\end{eqnarray}
\hrule
\end{figure*}
where $\Pr \left( {{\bm Z}\left( t \right) = {z_i}} \right) = {\pi _{{z_i}}}$, and ${\cal K}=\{1,2,\cdots, {N^2}+N +2\}$. Then we have an upper bound for the optimization problem \textbf{P1} \eqref{opt}, i.e, $V{O^*} \le {\phi ^*}$.

\textbf{Proof} of \emph{Lemma} 2:  For any stable policy, for any queue, the time average input rate cannot exceed the time average output rate. From constraints \eqref{EMenergy_availability},\eqref{EHenergy_availability},\eqref{EGenergy_availability},\eqref{Data_availability},we have that, for any queue, the input rate should be more than the output rate, so the time average input rate is equal to the time average output rate. By using Caratheodory's theorem (see, for example, \cite{Bertsekas2003}), and comparing the problem \textbf{P1} \eqref{opt} with \eqref{phi}, we conclude that $V{O^*}$ is one feasible solution to \eqref{phi}, so  $V{O^*} \le {\phi ^*}$. $\square$

Now, we begin to give the proof of  Part (B) in Theorem 2.
We can see from the problem \textbf{P2} that, after using Lyapunov optimization, our optimization problem is to minimize  ${\hat \Delta _V}\left( t \right)$. However, what we actually solve in the section IV is to maximize  $  \hat \Delta _V^o\left( t \right)+ \epsilon \sum\limits_{n \in {\cal N}} {\sum\limits_{f \in {\cal F}} {\sum\limits_{b \in {\cal O}\left( n \right)}\sum\limits_{s \in {\cal N}_s^f} {\sum\limits_{d \in {\cal N}_d^f}  } {x_{nb}^{fsd}\left( t \right)} } }$ by introducing  $\sigma $, where  $ \hat \Delta _V^o\left( t \right)$  is the function $ {\hat \Delta _V}\left( t \right)$ minimized under our proposed algorithm. Then, we have
\begin{eqnarray}
  \hat \Delta _V^o\left( t \right) &+& \epsilon \sum\limits_{n \in {\cal N}} {\sum\limits_{f \in {\cal F}} {\sum\limits_{b \in {\cal O}\left( n \right)}\sum\limits_{s \in {\cal N}_s^f} {\sum\limits_{d \in {\cal N}_d^f}  } {\tilde{x}_{nb}^{fsd,o}\left( t \right)} } }\notag\\
 &\le&  \hat \Delta _V^{F}\left( t \right) + \epsilon \sum\limits_{n \in {\cal N}} {\sum\limits_{f \in {\cal F}} {\sum\limits_{b \in {\cal O}\left( n \right)}\sum\limits_{s \in {\cal N}_s^f} {\sum\limits_{d \in {\cal N}_d^f}  } {\tilde{x}_{nb}^{fsd,F}\left( t \right)} } }\notag
\end{eqnarray}
Since $0 \le \epsilon \sum\limits_{n \in {\cal N}} {\sum\limits_{f \in {\cal F}} {\sum\limits_{b \in {\cal O}\left( n \right)}\sum\limits_{s \in {\cal N}_s^f} {\sum\limits_{d \in {\cal N}_d^f}  } {x_{nb}^{fsd}\left( t \right)} } } \le NFN_s N_d\epsilon l_{\max}X_{\max}$,
we have $ \hat \Delta _V^o\left( t \right) \le  {\hat \Delta _V^{F}}\left( t \right)+NFN_s N_d\epsilon l_{\max}X_{\max}$, where $F$ represents any other feasible policy. Compared to \eqref{phi}, we have that $-\mathbb{E} \left\{ { {{\hat \Delta }_V^{F}}\left( t \right)|{\bm Z}\left( t \right)} \right\}={\phi ^*}$. Then, we have
\begin{eqnarray}
\Delta \left( t \right)&-& V\mathbb{E}\left\{ {O\left( t \right)\left| {{\bm{Z}}\left( t \right)} \right.} \right\}\notag\\
&\le& B + \mathbb{E}\left\{ { \hat \Delta _V^o\left( t \right)\left| {{\bm{Z}}\left( t \right)} \right.} \right\}\nonumber\\
 &\le& B +\mathbb{E}\left\{ { {{\hat \Delta }_V^{F}}\left( t \right)\left| {{\bm{Z}}\left( t \right)} \right.} \right\}+N_s N_d\epsilon  l_{\max}X_{\max}\nonumber\\
 &=& \tilde{B} - {\phi ^*}
\end{eqnarray}
where $\tilde{B}=B+NFN_s N_d\epsilon  l_{\max}X_{\max}$.
 Based on the conclusion of \emph{Lemma} 1, i.e., $V{O^*} \le {\phi ^*}$, we have
\[\Delta \left( t \right) - V\mathbb{E}\left\{ {O\left( t \right)\left| {{\bm{Z}}\left( t \right)} \right.} \right\} \le \tilde{B}  - V{O^*}.\]
Then,
\begin{eqnarray}
&&\mathop {\lim }\limits_{T \to \infty } \frac{1}{T}\sum\limits_{t = 0}^{T - 1} \mathbb{E}{\left\{ {\Delta \left( t \right) - V\mathbb{E}\left\{ {O\left( t \right)\left| {{\bm{Z}}\left( t \right)} \right.} \right\}} \right\}}\nonumber \\
 &&= \mathop {\lim }\limits_{T \to \infty } \frac{1}{T}\sum\limits_{t = 0}^{T - 1} {\mathbb{E}\left\{ {L\left( {t + 1} \right) - L\left( t \right) - VO\left( t \right)} \right\}}\nonumber \\
&& = \mathop {\lim }\limits_{T \to \infty } \frac{1}{T}\mathbb{E}\left\{ {L\left( T \right) - L\left( 0 \right)} \right\} - \mathop {\lim }\limits_{T \to \infty } \frac{1}{T}\sum\limits_{t = 0}^{T - 1} {V\mathbb{E}\left\{ {O\left( t \right)} \right\}} \nonumber\\
&& \le \tilde{B} - V{O^*}\nonumber
\end{eqnarray}
Thus, we have
\begin{equation}\label{8}
 \mathop {\lim }\limits_{T \to \infty } \frac{1}{T}\sum\limits_{t = 0}^{T - 1} {\mathbb{E}\left\{ {O\left( t \right)} \right\}}  \ge {O^*} - \frac{{\tilde{B} }}{V},
\end{equation}
So, this completes the proof of \textbf{Theorem 2}(\textbf{B}). $\square$

\section{Proof of Part (\textbf C) in Theorem 2}
In order to prove \eqref{ea1}, we first assume that
$E_{n}^{H}\left( t \right) < P_{n,\max }^{Total}$, when node
$n \in {{\cal N}_H}$
allocates nonzero power for data sensing, compression and transmission.

\textbf{Case}1:When there exits energy consumption for data sensing/compression at node $n$, from \eqref{SR},if $E_{n}^{H}\left( t \right) < P_{n,\max }^{Total}$,
\begin{eqnarray}
 &&\sum\limits_{m:n \in \underline{{\cal N}}_{sm}^f} {\rho _m^{f*}\left( t \right)}- \sum\limits_{d \in {\cal N}_d^f} {Q_n^{fnd}(t)}  + A_n(t)\tilde P_f^S\notag\\
  &&\le \sum\limits_{m:n \in \underline{{\cal N}}_{sm}^f} {\rho _m^{f*}\left( t \right)}  + (E_n^H(t)-\theta_n^{eH})\tilde P_f^S\notag\\
  &&\le \beta V-{\max\{\frac{{1}}{\sigma},\delta BW\}} N_sN_d \beta V\tilde P_f^S \notag\\
  &&\le \beta V-\frac{{1}}{\sigma}N_sN_d \beta V\tilde P_f^S \le 0\notag
\end{eqnarray}
from \eqref{Anb} and $Q_n^{fsd}(t)\ge 0$ we have the first inequality, from \eqref{rhob} and \eqref{ehu} we have the second inequality, and from the definition of ${\sigma}$ we have the last inequality. The last inequality shows that if $E_{n}^{H}\left( t \right) < P_{n,\max }^{Total}$, the optimal solution of \eqref{SR} is zero, which means there exits no energy consumption for data sensing/compression.

\textbf{Case}2:It is easy to verify that the following inequation holds according to the definition of ${C_{ab}}\left( {{{\bm p}^T}\left( t \right),{\bm S}\left( t \right)} \right)$:
\begin{equation}
 {C_{ab}}\left( {{{\bm p}^T}\left( t \right),{\bm S}\left( t \right)} \right) \le {C_{ab}}\left( {{{\bm p}^{T'}}\left( t \right),{\bm S}\left( t \right)} \right)\label{as2}
\end{equation}
where
${{\bm p}^{T'}}\left( t \right)$ obtained by setting
$p_{{nm}}^T\left( t \right)$ of
${{\bm p}^T}\left( t \right)$ to zero, $\left( {a,b} \right) \in {\cal L}$ and $\left( {a,b} \right) \ne \left( {n,m} \right)$.

Whenever the link $(n,m)$ do transmission, for each session $f$ we get
\begin{flalign}\label{addPartC0}
 & W_{nm}^f\left( t \right) &\notag\\
 &={\left[ {\sum\limits_{s \in {\cal N}_s^f} {\sum\limits_{d \in {\cal N}_d^f} {(Q_n^{fsd}(t) - Q_b^{fsd}(t)-\epsilon)} }+A_{m}(t) \tilde P_m^R } \right]^ + }&\notag\\
 &\le {\left[ \sum\limits_{s \in {\cal N}_s^f} \sum\limits_{d \in {\cal N}_d^f} {(Q_n^{fsd}(t)-\epsilon)}  \right]^ + }&
\end{flalign}

By plugging the definition of $\epsilon$ and \eqref{Qbound} into \eqref{addPartC0}, we have
\begin{flalign}\label{addPartC1p}
  &W_{nm}^f\left( t \right) &\notag\\
  &\le \sum\limits_{s \in {\cal N}_s^f} \sum\limits_{d \in {\cal N}_d^f} {\left[{\beta V + R_{\max } - {l_{\max }}{X_{\max }} - R_{\max }} \right]^ + }&\notag\\
  &=N_sN_d{\left[ {\beta V - {l_{\max }}{X_{\max }}} \right]^ + }&
\end{flalign}
Moreover, we have
\begin{eqnarray}\label{addPartC1}
 W_{nm}^{*}(t)&\le& N_sN_d{\left[ {\beta V - {l_{\max }}{X_{\max }}} \right]^ + }
\end{eqnarray}
We assume that
$E_{n}^{H}\left( t \right) < P_{n,\max }^{Total}$, when node
$n \in {{\cal N}}$
allocates nonzero power for data transmission.
Furthermore, we assume that the power allocation control vector ${{\bm p}^{T*}}(t)$ is the optimal solution to \eqref{opeq}, and without loss of generality, there exists some $p_{{nm}}^{{T^{\rm{*}}}}(t) > 0$. By setting $p_{{nm}}^{T*} (t)= 0$ in ${{\bm p}^{T*}}(t)$, we get another power allocation control vector ${{\bm p}^T}(t)$.
We denote $G\left( {{{\bm p}^T}(t),{\bm S}(t)} \right)$ as the objective function of \eqref{opeq}. 
In this way, we get:
\begin{eqnarray}
&&G\left( {{{\bm p}^{T*}(t)},{\bm S}(t)} \right) - G\left( {{{\bm p}^T}(t),{\bm S}(t)} \right)\\
&&= \sum\limits_{n \in {\cal N}} \sum\limits_{b \in {\cal O}\left( n \right)} [ {C_{nb}}({{\bm p}^{T*}(t)},{\bm S}(t))  \notag\\
&& - {C_{nb}}({{\bm p}^T}(t),{\bm S}(t)) ]\tilde W_{{nb}}^*\left( t \right)
+ \left ({E_{n}^H\left( t \right) - \theta _{n}^{eH}}\right )p_{{nm}}^{T*}\notag
\end{eqnarray}
From \eqref{as2}, we have
${C_{nb}}({{\bm p}^{T*}(t)},{\bm S}(t)) - {C_{nb}}({{\bm p}^T(t)},{\bm S}(t)) \le 0$ for
$b \ne m$. So
\begin{eqnarray}\label{addPartC3}
&&G\left( {{{\bm p}^{T*}(t)},{\bm S}(t)} \right) - G\left( {{{\bm p}^T(t)},{\bm S}(t)} \right)\\
&&\le {C_{nm}}({{\bm p}^{T*}(t)},{\bm S}(t))\tilde W_{{nb}}^*\left( t \right) { + \left ({E_{n}^H\left( t \right) - \theta _{n}^{eH}}\right )p_{{nm}}^{T*}}\notag
\end{eqnarray}
According to our assumption $E_{n}^H\left( t \right) < P_{n,\max }^{Total}$ and the definition of $\theta_n^{eH}$ in \eqref{thetaeh}, we have
\begin{eqnarray}\label{addPartC2}
  E_{n}^H\left( t \right) - \theta _n^{eH} &<& P_{n,\max }^{Total} - \theta _n^{eH}\nonumber\\
  &=&-{\max\{\frac{{1}}{\sigma},\delta BW\}} N_sN_d \beta V
\end{eqnarray}
Plugging \eqref{PowerFeature1}, \eqref{addPartC1} and \eqref{addPartC2} into \eqref{addPartC3}, we have
\begin{eqnarray}
&&G\left( {{{\bm p}^{T*}},{\bm S}} \right) - G\left( {{{\bm p}^T},{\bm S}} \right)\notag\\
&&\le \delta BWp_{nm}^{T*} N_sN_d{\left[ {\beta V - {l_{\max }}{X_{\max }}} \right]^ + }\notag\\
&&-{\max\{\frac{{1}}{\sigma},\delta BW\}} N_sN_d \beta Vp_{{nm}}^{T*} < 0\nonumber
\end{eqnarray}
the last inequality is due to the assumption $ {\max\{\frac{{1}}{\sigma},\delta BW\}} \ge \delta BW$.
From the last inequality, we can see that if $E_{n}^H\left( t \right) < P_{n,\max }^{Total}$, ${{\bm p}^{T*}}$  is not the optimal solution to \eqref{opeq}.

\textbf{Case}3:When node $n$ receives data from other node $a \in {\cal I}(n)$ in the network, if $E_{n}^H\left( t \right) < P_{n,\max }^{Total}$,for each session $f$ we get
\begin{flalign}\label{addPartC0}
  &W_{an}^f\left( t \right) &\notag\\
  &={\left[ {\sum\limits_{s \in {\cal N}_s^f} {\sum\limits_{d \in {\cal N}_d^f} {(Q_a^{fsd}(t) - Q_n^{fsd}(t)-\epsilon)} }+A_{n}(t) \tilde P_n^R } \right]^ + }&\notag\\
 &\le {\left[ \sum\limits_{s \in {\cal N}_s^f} \sum\limits_{d \in {\cal N}_d^f} Q_n^{fsd}(t) +(E_n^H(t)-\theta_n^{eH})\tilde P_n^R \right]^ + }&\notag\\
 &\le{\left[ N_sN_d\beta V -{\max\{\frac{{1}}{\sigma},\delta BW\}} N_sN_d \beta VP_n^R \right]^ + }\le0&\notag
\end{flalign}
from the definition of $\epsilon$, \eqref{Anb}and $Q_n^{fsd}(t)\ge 0$, we have the first inequality.From \eqref{Qbound} and \eqref{ehu}, we have the second inequality, and the last inequality is due to the definition of $\sigma$. The last inequality shows that if $E_{n}^H\left( t \right) < P_{n,\max }^{Total}$, the node $n$ can't receive the endogenous data.

Above all, if $E_{n}^H\left( t \right) \geq P_{n,\max }^{Total}$, the node $n$ can't do any work such as data sensing/compression, transmission and acceptation, which is opposite with our assumption. So, $E_{n}^H\left( t \right) \geq P_{n,\max }^{Total}$, which completes the proof of \eqref{ea1}.

When the EG node is discharging, from \eqref{EPBCD}, we obtain the following:
 $ E_n^G\left( t \right) \ge \theta _n^G-{\lambda _n} \le \frac{{1}}{{\sigma}}\beta V +d_n^{\max}-\frac{{1}}{{\sigma}}\beta V =d_n^{\max}$, where the first inequality is obtained from \eqref{thetaeg} and \eqref{lambdab}, the \eqref{ea2} is proved.

 The prove of \eqref{ea3} is similar with \eqref{ea2}, therein, we omit it for brief.

Until now, we complete the proof of \textbf{Theorem 2}(\textbf{C}). $\square$

\section{Proof of Part (\textbf D) in Theorem 2}

For each node $n \in {{\cal N}}$, when any data of the $f$-th session from source $s$ to sink $d$ is transmitted to other node, we can get from the session scheduling component that $Q_n^{fsd}(t) - Q_b^{fsd}(t)+ {A_b}\left( t \right)\tilde P_b^R - \epsilon>0$, i.e.,
$Q_n^{fsd}(t) > Q_b^{fsd}(t)- {A_b}\left( t \right)\tilde P_b^R + \epsilon>0$
From the definition of $\epsilon$ ,$Q_b^{fsd}(t)\ge 0$ and ${A_b}\left( t \right) \le 0$
we have $Q_n^f(t) >  {l_{\max }}{X_{\max }} + {R_{\max }}$, which completes the proof of \textbf{Theorem 2}(\textbf{D}). $\square$

\section{Proof of Lemma 1}

We assume $\bm \lambda^{*}$ and $\bm \rho^{*}$ are the optimal solution of \eqref{opt3}. For \eqref{D}, let $\bm D^{*}$ be the optimal solution, hence, we have
\begin{eqnarray}
&&V{\varpi _1}U_n^f\left( {D_n^{f{\rm{*}}}(t)} \right) + \log \left( {D_n^{f{\rm{*}}}(t)} \right)\sum\limits_{m:n \in \underline{\cal N}_{sm}^f} {\rho _m^{f{\rm{*}}}} \left( t \right)\notag\\
&& \ge V{\varpi _1}U_n^f\left( {{D_{\max }}} \right) + \log \left( {{D_{\max }}} \right)\sum\limits_{m:n \in \underline{\cal N}_{sm}^f} {\rho _m^{f{\rm{*}}}} \left( t \right)\notag
\end{eqnarray}
Next, we have
\begin{eqnarray}
&& \sum\limits_{m:n \in \underline{{\cal N}}_{sm}^{f}} {\rho _m^{f*}\left( t \right)}\le V\varpi_1
\frac{{U_n^f\left( {D_n^{f*}\left( t \right)} \right) - U_n^f\left( {{D_{\max }}} \right)}}{{\log \left( {{D_{\max }}} \right) - \log \left( {D_n^{f*}\left( t \right)} \right)}}\notag\\
&&\le V\varpi_1
{\sup _{{D_{\min }} \le D_n^f\left( t \right) \le {D_{\max }}}}\frac{{U_n^f\left( {D_n^f\left( t \right)} \right) - U_n^f\left( {{D_{\max }}} \right)}}{{\log \left( {{D_{\max }}} \right) - \log \left( {D_n^f\left( t \right)} \right)}}\notag
\end{eqnarray}

Similarly, for \eqref{SR} at node $n \in \cal N_G \cup \cal N_M$, let $\bm r^{*}$ be the optimal solution, we have
$\sum\limits_{m:n \in \underline{{\cal N}}_{sm}^f} {\rho _m^{f*}\left( t \right)}- \sum\limits_{d \in {\cal N}_d^f} {Q_n^{fnd}(t)}  - \lambda_n^{*}\tilde P_f^S \ge 0$,
thus
\begin{eqnarray}
 \lambda_n^{*} \tilde P_f^S &\le& \sum\limits_{m:n \in \underline{{\cal N}}_{sm}^f} {\rho _m^{f*}\left( t \right)}- \sum\limits_{d \in {\cal N}_d^f} {Q_n^{fnd}(t)} \notag\\
  &&\le \sum\limits_{m:n \in \underline{{\cal N}}_{sm}^f} {\rho _m^{f*}\left( t \right)}\le \beta V\notag
\end{eqnarray}
and from the definition of $\sigma$, we have
$\lambda_n^{*}\sigma \le \beta V$, i.e.,$\lambda_n^{*}\le \frac{{1}}{{\sigma}}\beta V$.

Above all, we complete the proof of Lemma 1.

\end{document}